\documentclass[reprint, superscriptaddress, amsmath, amssymb, aps, prl]{revtex4-2}

\usepackage{braket}
\usepackage{graphicx}
\graphicspath{{Images/}}

\usepackage[dvipsnames,table]{xcolor}
\usepackage[colorlinks=true,linkcolor=RoyalBlue,citecolor=Brown,urlcolor=Black,filecolor=Brown]{hyperref}

\newcommand{\PRLsection}[1]{\textit{#1} ---}

\usepackage{booktabs,todonotes,upgreek,soul}

\makeatletter
\def\CT@@do@color{%
  \global\let\CT@do@color\relax
        \@tempdima\wd\z@
        \advance\@tempdima\@tempdimb
        \advance\@tempdima\@tempdimc
\advance\@tempdimb\tabcolsep
\advance\@tempdimc\tabcolsep
\advance\@tempdima2\tabcolsep
        \kern-\@tempdimb
        \leaders\vrule
                \hskip\@tempdima\@plus  1fill
        \kern-\@tempdimc
        \hskip-\wd\z@ \@plus -1fill }
\makeatother

\begin{document}

\newcommand{\thf}[1]{$^{#1}\mathrm{Th}\mathrm{F}^+$}

\title{Nuclear Schiff moment of the fluorine isotope $^{19}$F}

\author{Kia Boon Ng}
\email{kbng@triumf.ca}
\affiliation{TRIUMF, Vancouver, British Columbia, V6T 2A3, Canada}

\author{Stephan Foster}
\affiliation{TRIUMF, Vancouver, British Columbia, V6T 2A3, Canada}
\affiliation{Department of Engineering Physics, McMaster University, Hamilton, Ontario, L8S 4M1, Canada}

\author{Lan Cheng}
\affiliation{Department of Chemistry, The Johns Hopkins University, Baltimore, MD 21218, USA}

\author{Petr Navr\'atil}
\affiliation{TRIUMF, Vancouver, British Columbia, V6T 2A3, Canada}

\author{Stephan Malbrunot-Ettenauer}
\affiliation{TRIUMF, Vancouver, British Columbia, V6T 2A3, Canada}
\affiliation{Department of Physics, University of Toronto, Toronto, Ontario, M5S 1A7, Canada}

\date{\today}

\begin{abstract}
Nuclear Schiff moments (NSMs) are sensitive probes for physics beyond the Standard Model of particle physics, signaling violations of time-reversal and parity-inversion symmetries in atomic nuclei. In this Letter, we report the first-ever calculation of a NSM in a nuclear \textit{ab initio} framework, employing the no-core shell model to study the fluorine isotope $^{19}$F. We further perform quantum-chemistry calculations to evaluate the sensitivity of the hafnium monofluoride cation, HfF$^+$, to the NSM of $^{19}$F. Combined with recent high-precision measurements of the molecular electric dipole moment of HfF$^+$ \cite{Roussy2023an}, our results enable the first experimental bound on the NSM of $^{19}$F. Although the resulting bounds on the pion-nucleon-nucleon ($\pi$NN) coupling constants are not yet the most stringent, this work establishes the foundation for constraining $\pi$NN interactions using nuclear \textit{ab initio} methods.
\end{abstract}

\maketitle

\PRLsection{Introduction}
    Rapid advancements in the control and manipulation of individual quantum states in atoms and molecules have supercharged progress in the fields of quantum information, quantum simulation, and the probing of chemical reactions at the quantum level \cite{langen2024quantum, demille2024quantum, ye2024essay}. Over the years, numerous experiments have leveraged these quantum-enabled technologies to probe for electric dipole moments (EDMs) of elementary and composite particles \cite{Roussy2023an, acme2018improved, hudson2011improved, eckel2013search, zheng2022measurement, Sachdeva2019new, Graner2016reduced, bishof2016improved, regan2002new, murthy1989new, Cho1991search, nedm2020}. Measurements of EDMs can help constrain the nature of new physics \cite{POSPELOV2005119, ENGEL201321, Chupp2019electric, cesarotti2019interpreting, Ramsey1950on} and guide us to a better understanding of the underlying mechanisms of the universe.

    One example of an EDM sensitive to fundamental symmetry violations is the molecular EDM induced by a nuclear Schiff moment (NSM) of a constituent atom, e.g., $^{19}$F in HfF$^+$. The physics of the NSM has been covered comprehensively in literature, e.g., Refs.\ \cite{engel2025nuclear, flambaum2020electric, dobaczewski2018correlating, engel2000nuclear, spevak1997enhanced, auerbach1996collective}. 
    In brief, the NSM arises from nuclear electric moments that are only partially screened by the surrounding electron cloud due to relativistic effects and the finite size of the nucleus \cite{schiff1963measurability, flambaum2012extension}. Parity-inversion ($\mathcal{P}$) and time-reversal ($\mathcal{T}$) violation within the atomic nucleus can generate a NSM, which in turn induces a measurable EDM in atoms and molecules. While $\mathcal{P,T}$-violating effects predicted by the Standard Model produce EDMs well below current experimental sensitivity, many beyond-Standard-Model scenarios predict enhancements potentially within reach of ongoing or near-future measurements. As such, EDM experiments provide powerful probes of new $\mathcal{P,T}$-violating physics.
    
    A key component of interpreting EDM measurement is the theoretical connection between experimental observables and fundamental $\mathcal{P,T}$-violating phenomena. This requires two key calculations: the molecular sensitivity coefficient ($W_S$) which relates the observed EDM to the NSM ($S$), and a nuclear structure calculation that connects the NSM to the underlying $\mathcal{P,T}$-violating physics. The energy shift, $\Delta E_\mathrm{NSM}$, of a molecular state due to the NSM of a constituent nucleus is given by \cite{flambaum2020electric}:
    \begin{equation}\label{eq:WSS} 
        \Delta E_\mathrm{NSM} = W_S \, S \, \left\langle \frac{\mathbf{I}\cdot \hat{n}}{I} \right\rangle, 
    \end{equation}
    where $\mathbf{I}$ is its nuclear spin vector, $I$ is its magnitude, and $\hat{n}$ is the unit vector along the internuclear axis. The expectation value is evaluated for the specific EDM-sensitive state utilized in the experiment.

    State-of-the-art relativistic quantum chemistry methods, using \textit{ab initio} approaches and extended basis sets, can determine molecular sensitivity factors with high accuracy and uncertainties at the sub-10\% level~\cite{Gaul20,Skripnikov20,Abe20,marc2023candidate,chen2024relativistic}.
    
    In contrast, solving the nuclear many-body problem with comparable precision remains a major theoretical challenge, particularly when linking nuclear structure to searches for new physics. Traditionally, this connection has relied on phenomenological nuclear models that, while powerful, are typically finely tuned and model-dependent. A striking example is the calculation of nuclear matrix elements for neutrinoless double-beta decay, where model predictions can differ by more than a factor of three \cite{Engel2017}. Similar discrepancies are observed in phenomenological estimates of NSMs, which can vary by an order of magnitude and even disagree in sign~\cite{engel2025nuclear}, underscoring the need for more systematic and predictive nuclear theory.
    
    Over the past decade, nuclear structure theory has advanced significantly, driven by improvements in many-body methods, increased computational power, and more accurate descriptions of nuclear forces. In particular, chiral effective field theory ($\chi$EFT) provides a systematic framework that connects the symmetries of quantum chromodynamics to low-energy nuclear interactions, enabling controlled expansions of nuclear forces at the scale where protons and neutrons form atomic nuclei \cite{RevModPhys.81.1773,Machleidt2011,RevModPhys.92.025004}. These interactions are now routinely used in nuclear \textit{ab initio} calculations, offering high precision and reliable uncertainty quantification \cite{Hergert2020}. Extending \textit{ab initio} methods to the calculation of NSMs thus holds promise for improving the theoretical reliability of NSM estimates, analogous to recent progress in \textit{ab initio} studies of neutrinoless double-beta decay \cite{PhysRevLett.124.232501,PhysRevLett.126.182502,PhysRevLett.126.042502,PhysRevLett.132.182502}.
    
    In this Letter, we present the first nuclear \textit{ab initio} calculation of a NSM, focusing on the fluorine isotope $^{19}$F. This nucleus is ideally suited for several reasons. First, its relatively low mass makes it tractable within the no-core shell model (NCSM), which treats all 9 protons and 10 neutrons explicitly. Hence, it will serve as a reliable benchmark for future \textit{ab initio} efforts aimed at calculating NSMs in heavier, octupole-deformed systems. Second, $^{19}$F exhibits a low-lying opposite-parity excited state ($I^\pi = 1/2^-$) just $\approx 110~\mathrm{keV}$ above its $1/2^+$ ground state~\cite{Tilley1995}, which strongly amplifies NSM-induced effects~\cite{flambaum2020electric}. Finally, we show that the recent high-precision measurement of the molecular EDM in HfF$^+$~\cite{Roussy2023an}, while primarily constraining the electron EDM, can also be interpreted as a stringent limit on the NSM of $^{19}$F. To support this, we also perform quantum-chemistry calculations of the molecular sensitivity factor for HfF$^+$ and other fluorine-containing molecules. While the resulting bounds on the pion-nucleon-nucleon ($\pi$NN) coupling constants are not yet the most stringent, this work lays the groundwork for future $\pi$NN constraints that will be derived from nuclear \textit{ab initio} calculations.
    
\vspace{2mm}
\PRLsection{Nuclear Schiff moment of $^{19}$F}
    We perform nuclear \textit{ab initio} calculations of the $^{19}$F NSM using the NCSM approach \cite{BARRETT2013131} in combination with the Lanczos strength method \cite{Haydock1974, Marchisio2003} that was previously applied in NCSM evaluations of anapole and electric dipole moments of light nuclei \cite{Hao2020, froese2021ab}. As input, we use primarily two sets of parity-conserving (PC) $\chi$EFT nucleon-nucleon (NN) and three-nucleon (3N) interactions: NN-N$^4$LO~\cite{Entem2017} + 3N$^*_\mathrm{lnl}$~\cite{Kravvaris2023,Jokiniemi2024,Girlanda2011} and NN-N$^3$LO~\cite{Entem2003} + 3N$_\mathrm{lnl}$~\cite{Soma2020}.
    The leading-order (LO) $\mathcal{P,T}$-violating NN interaction due to the one-pion exchange is taken in the form introduced in Ref.~\cite{Liu2004}, which is equivalent to the LO $\chi$EFT $\mathcal{P,T}$-violating NN interaction~\cite{deVries2020}. It should be noted that contact interactions can also play an important role: for CP-violating sources that preserve chiral symmetry, they enter at the same order in chiral effective field theory as the pion-exchange contribution~\cite{deVries2020}. We defer a quantitative assessment of their impact on the Schiff moments of light nuclei to a separate work. 

    The nuclear Schiff moment operator associated with the charge distribution is given by~\cite{engel2025nuclear}:
    \begin{equation}\label{Sch_op}
    \mathbf{S} = \frac{e}{10} \sum_{p=1}^{Z} \left( r_p^2 - \frac{5}{3} \langle r^2 \rangle_{\text{ch}} \right) \mathbf{r}_p \; ,
    \end{equation}
    where $e$ is the elementary charge, $\langle r^2 \rangle_{\text{ch}}$ the charge radius of the nuclear ground state, $Z$ is the proton number, $\mathbf{r}_p$ is the position vector of each proton $p$ relative to the center-of-mass of the nucleus, and $r_p$ is its magnitude. In addition to this charge distribution term, a separate ``nucleon component'' of the Schiff moment arises from the intrinsic EDMs of the nucleons. This contribution is neglected here, as the $\mathcal{P,T}$-violating NN interaction is expected to dominate \cite{engel2025nuclear,ENGEL201321}. 
    
    The Schiff moments are evaluated by computing the matrix elements:
    \begin{equation}\label{eq:D2}
      {S}=\bra{A , {\rm g.s.} , I^\pi, I_z{=}I} {S}_z \ket{A , {\rm g.s.} , I,  I_z{=}I} 
                       + {\rm h.c.},
    \end{equation}
    where ``g.s.'' indicates the nuclear ground state, $I$ is the nuclear spin quantum number, $\pi$ is the parity, $I_z$ and $S_z$ are the $z$ component of the nuclear spin vector and the Schiff moment operator, respective, with the $z$ direction defined to be along the intrinsic body axis, and $A$ encapsulates all other relevant quantum numbers.
    We solve the standard Schr\"{o}dinger equation using the PC Hamiltonian and obtain the $\ket{A , {\rm g.s.} , I^\pi}$ wave function. We then invert the generalized Schr\"{o}dinger equation with an inhomogeneous term,
    \begin{equation}\label{inhomeq}
      (E_{\rm g.s.}^{I^\pi}-H)  |A , {\rm g.s.} , I \rangle = V_{\rm NN}^{\rm PTV}|A , {\rm g.s.} , I^\pi \rangle \; ,
    \end{equation}
    where $H$ is the PC Hamiltonian, $E_{\rm g.s.}^{I^\pi}$ is the energy of the pure-parity nuclear ground state $\ket{A , {\rm g.s.} , I^\pi}$, and $|A , {\rm g.s.} , I \rangle$ is the ground state wave function that includes the unnatural parity admixture due to the $\mathcal{P,T}$-violating interaction, $V_{\rm NN}^{\rm PTV}$~\cite{froese2021ab}. See also Equation~(S1) in Supplemental Material~\cite{SM}. Note that we shift the calculated $E_{\rm g.s.}^{I^\pi}$ energy to match experimental excitation energy of the lowest opposite-parity state (here $1/2^-_1$).
        
        Our results yield the following expression for the $^{19}$F NSM: 
        \begin{equation}\label{eq:S19F} 
            S(^{19}\mathrm{F}) = (-2.9\, g\bar{g}_0 - 2.4\, g\bar{g}_1 - 2.0\, g\bar{g}_2) \times 10^{-2}~e~\mathrm{fm}^3,
        \end{equation}
        where $g$ denotes the $\mathcal{P,T}$-conserving pion-nucleon-nucleon ($\pi$NN) coupling constant; and $\bar{g}_0$, $\bar{g}_1$, and $\bar{g}_2$ are the isoscalar, isovector, and isotensor $\mathcal{P,T}$-violating $\pi$NN coupling constants, respectively. A nonzero value of any of the couplings $\bar{g}_{0,1,2}$ would indicate physics beyond the Standard Model. By varying the NCSM basis size, the harmonic oscillator frequency, and the $\chi$EFT PC interactions, we estimate the uncertainty of the results given in Equation~\eqref{eq:S19F} at about 50\%. The calculation details are given in the Supplemental Material~\cite{SM}. %

        The NSM of $^{19}$F is significantly enhanced compared to other light nuclei~\cite{Foster2025} due to the presence of a relatively low-lying opposite-parity partner of its ground state and a large Schiff operator matrix element between the two states. Remarkably, its magnitude is comparable to that of the much heavier nucleus $^{129}$Xe, as obtained from phenomenological large-scale shell model calculations~\cite{yanase2020large}. However, the lighter mass of $^{19}$F results in smaller coefficients for the $\pi$NN coupling terms than those in heavier and octupole-deformed nuclei such as $^{225}$Ra~\cite{dobaczewski2005nuclear} and $^{227}$Ac~\cite{athanasakiskaklamanakis2025}.

        Nevertheless, the light mass of $^{19}$F enables its NSM to be computed using \textit{ab initio} methods that provide a more detailed and reliable description of the nuclear structure than approaches typically used for heavier nuclei \cite{sushkov1984possibility, flambaum1986p, yoshinaga2018nuclear, yoshinaga2020schiff, yanase2020large, dmitriev2003p, dmitriev2005effects, ban2010fully, jesus2005time, dobaczewski2005nuclear, dobaczewski2018correlating, spevak1997enhanced}. Applying the NCSM with the Lanczos strength method, we are able to compute exactly the nuclear many-body Green's function in Equation~\eqref{inhomeq} and take into account the effect of all intermediate states the $\mathcal{P,T}$-violating NN interaction and the Schiff operator connect to. We can then assess the contribution of the lowest opposite-parity partner, which is typically the only contribution considered in heavy systems. We find that the NSM of $^{19}$F is dominated by this contribution. Interestingly, this is not the case for the nuclear EDM of $^{19}$F where the lowest opposite-parity partner contribution is negligible. The reason for that is a differing structure of the two states: The $1/2^+$ ground state is a shell-model like state with large $S$-wave $^{18}$O+p amplitude while the $1/2^-_1$ state exhibits $\alpha$-clustering with large $S$-wave $^{15}$N+$\alpha$ amplitude and a negligible $P$-wave $^{18}$O+p amplitude. Consequently, the matrix elements of the $E1$ operator and of the second term of the Schiff operator in Equation~\eqref{Sch_op} ($\varpropto \mathbf{r}$) are very small while that of the first term of the Schiff operator ($\varpropto r^2 \mathbf{r}$) is enhanced, see the Supplemental Material~\cite{SM} for further details.
        
\vspace{2mm}
\PRLsection{Molecular sensitivity factors for $^{19}$F-containing molecules}
        According to Equation~\eqref{eq:WSS}, the NSM-induced energy shift in a molecule is a product of the NSM and a molecular sensitivity factor proportional to the gradient of the electron density at the targeted nucleus \cite{flambaum2020electric}. The electron cloud around a $^{19}$F nucleus in a diatomic metal-fluoride molecule is heavily polarized due to the ionic metal-fluorine bond, enhancing the molecule’s sensitivity to $^{19}$F NSM measurements. We perform wavefunction-based {\it{ab initio}} electronic structure calculations for the $^{19}$F NSM molecular sensitivity factors for various molecules of interest. The results are shown in Table~\ref{tab:WS}. 
        \begin{table}[b]
           \caption{\textbf{Computed molecular sensitivity factors to the nuclear Schiff moment of $^{19}$F for select molecules.} $W_S$ is expressed in units of $\frac{e}{4\pi \epsilon_0 a_0^4} \approx 44.3~h~\mathrm{Hz}/(e~\mathrm{fm}^3)$.}
            \label{tab:WS}
            \begingroup
            \setlength{\tabcolsep}{4pt}
            \centering
            \rowcolors{2}{gray!20}{white}
            \begin{tabular}{l l l}
                \toprule
                Molecule & Molecular state & $W_S \left( \frac{e}{4\pi \epsilon_0 a_0^4} \right)$ \\
                \midrule
                HfF$^+$ & \hspace{3pt}$a\,^3\Delta_1$ & 115 \\ 
                ThF$^+$ & $X\,^3\Delta_1$ & \hphantom{1}99 \\ 
                SrF & $X\,^2\Sigma^+$ & \hphantom{1}52 \\
                BaF & $X\,^2\Sigma^+$ & \hphantom{1}48 \\
                YbF & $X\,^2\Sigma^+$ & \hphantom{1}59 \\
                TlF & $X\,^1\Sigma^+$ & \hphantom{1}74 \\
                RaF & $X\,^2\Sigma^+$ & \hphantom{1}47 \\
                \bottomrule
            \end{tabular}
            \endgroup
        \end{table}
        
        Major challenges in calculations of NSM molecular sensitivity factors include the high computational costs of relativistic electron-correlation calculations, the numerical stability issue in the differentiation of electronic energy with respect to NSM involving a very local interaction operator, and the need to go beyond standard basis sets to describe the local density variation. 
        We adopted our recent efficient implementation of relativistic exact two-component coupled-cluster singles and doubles (X2C-CCSD) \cite{Liu18,Liu18b,Zhang22} method that reduces the storage requirement by an order of magnitude and computing time by a factor of four. 
        We employ analytic X2C-CCSD gradient techniques \cite{Liu21} to avoid tedious and expensive numerical differentiation procedures. Accurate description for the electron density distribution in the immediate vicinity of the $^{19}$F nucleus is beyond the ability of standard F basis sets for molecular electronic structure calculations. We construct an extended basis set for $^{19}$F using the scheme described in Ref.~\cite{chen2024relativistic}. 
        The resulting ETB0 (30s30p4d3f2g) basis set for $^{19}$F provides sufficient flexibility to probe the electron-density gradient at the $^{19}$F nucleus. 
        
        Based on a systematic study of relativistic, correlation, and basis-set effects, we expect the errors for the X2C-CCSD $^{19}$F NSM molecular sensitivity factors in Table~\ref{tab:WS} to be well below 10\%. For example, the correction to $W_S$ of HfF$^+$ from the triple excitations amounts to less than 2\% of the total value. The inclusion of additional basis functions changes the computed value by less than 1\%. The difference between X2C and four-component results is less than 1\%. This error estimate is consistent with that in benchmark calculations reported in Ref.~\cite{chen2024relativistic}. More details can be found in the Supplemental Material~\cite{SM}.

        Of note, the magnitude of electron density close to the $^{19}$F nucleus is significantly smaller than the nuclei of heavy metals in these molecules. Therefore, the $^{19}$F NSM molecular sensitivity parameters are a few orders of magnitude smaller than those for the heavy atoms.

\vspace{2mm}
\PRLsection{The molecular EDM measurement in HfF$^+$ and its NSM interpretation} 
    Experimental EDM measurements of $^{19}$F-containing molecules have been reported for thallium fluoride (TlF) \cite{coveney1983parity,Cho1991search}, ytterbium fluoride (YbF) \cite{hudson2002measurement,hudson2011improved}, and HfF$^+$ \cite{cairncross2017precision, Roussy2023an}. While all of them are sensitive to the NSM of $^{19}$F, the HfF$^+$ result represents the most precise such measurement to date and its associated molecular sensitivity is the largest, see again Table~\ref{tab:WS}. We therefore employ the HfF$^+$ result to set the most stringent experimental upper bound on $^{19}$F's NSM. Note that the hafnium isotope employed in the HfF$^+$ experiment, $^{180}$Hf, possesses a nuclear spin of zero ($I = 0$). Thus, the measurement is insensitive to nuclear-spin-dependent contributions from hafnium.

    The HfF$^+$ EDM measurements were performed on the $a\,^3\Delta_1 (v=0, J=1, F=3/2, m_F=\pm 3/2)$ states, where $v$, $J$, and $F$ correspond to the vibrational, rotational, and hyperfine quantum numbers, respectively, and $m_F$ is the projection of $F$ onto the quantization axis. This manifold was chosen to maximize the molecule’s sensitivity to the electron’s EDM. In the more precise of the two HfF$^+$ measurements \cite{Roussy2023an}, the energy splitting between EDM-sensitive molecular states was determined to be
        \begin{equation}\label{eq:GenII} 
            h f = (-14.6 \pm 22.8_\mathrm{stat} \pm 6.9_\mathrm{syst})~h~\upmu\mathrm{Hz}, 
        \end{equation}
    where $h$ denotes Planck’s constant. This result was subsequently used to constrain the electron's EDM and the scalar–pseudoscalar nucleon–electron coupling, as reported in Ref.~\cite{Roussy2023an}. In addition to these two known sources of a molecular EDM in $^{180}$HfF$^+$, there exists a previously unaccounted-for contribution: the NSM of $^{19}$F. 

    To relate the HfF$^+$ EDM measurement to the NSM of $^{19}$F and the underlying $\mathcal{P,T}$-violating interactions, we make use of Equation~\eqref{eq:WSS}, informed by our \textit{ab initio} quantum chemistry and nuclear structure calculations, discussed above. In this case, the final term on the right-hand side in Equation~\eqref{eq:WSS} evaluates to $\left|\left\langle \frac{\mathbf{I}\cdot \hat{n}}{I} \right\rangle \right| = 1/2$ in the $^3\Delta_1 (v=0, J=1, F=3/2, m_F=\pm 3/2)$ manifold \cite{brown2003rotational}. Moreover, the experimental scheme measures the energy difference between the $m_F = +3/2$ and $m_F = -3/2$ states, in which the fluorine nuclear spin is oppositely oriented. A nonzero NSM shifts the energy of each state in opposite directions, so that the observed energy splitting corresponds to twice the NSM-induced shift in a single EDM-sensitive state, i.e., 
    \begin{equation}
        hf = 2 |\Delta E_\mathrm{NSM}|. 
    \end{equation}

    Combining these considerations with our theoretical results, specifically the molecular sensitivity factor $W_S$ in Table\,\ref{tab:WS} and the NSM expression given in Equation~\eqref{eq:S19F}, the measured frequency can be related, to two significant figures, 
    to the underlying $\mathcal{P,T}$-violating $\pi$NN coupling constants via:
    \begin{equation}\label{eq:S19F_number}
        h f = \left( -150 \, g \bar{g}_0 - 120 \, g \bar{g}_1 - 100 \, g \bar{g}_2 \right) \, h~\mathrm{Hz}.
    \end{equation} 

\vspace{2mm}
\PRLsection{Electric dipole moment of the fluorine nucleus} 
    In ions, the atomic nuclei are not completely shielded from any applied external electric field \cite{flambaum2012extension}. Hence, in addition to the energy shift induced by the NSM, the nuclear EDM, here that of $^{19}$F, produces shifts in the energy splitting between molecular states of opposite nuclear-EDM orientation with respect to the unscreened electric field, as described below: 
    \begin{equation}\label{eq:hfEDM}
        h f_\mathrm{EDM} = 2 \, |\mathbf{d} \cdot \pmb{\mathcal{E}}_\mathrm{unsc.}|,
    \end{equation}
    where $\mathbf{d}$ is the nuclear EDM and $\pmb{\mathcal{E}}_\mathrm{unsc.}$ is the unscreened electric field seen by the $^{19}$F atomic nucleus:
    \begin{equation}
        \pmb{\mathcal{E}}_\mathrm{unsc.} = \frac{M}{M_\mathrm{mol.}} \frac{Q_\mathrm{mol.}}{Z e} \, \pmb{\mathcal{E}}_\mathrm{ext.},
    \end{equation}
    where $M$ is the mass of the $^{19}$F atomic nucleus, $M_\mathrm{mol.}$ is the total mass of the molecular ion, $Q_\mathrm{mol.}$ is the net charge of the molecular ion, $Z$ is the number of protons in the $^{19}$F atomic nucleus, $e$ is the elementary charge, and $\pmb{\mathcal{E}}_\mathrm{ext.}$ is the applied external electric field.

    The nuclear EDM of $^{19}$F has been calculated using the NN-N$^{3}$LO + 3N$_\mathrm{lnl}$ interaction to be \cite{froese2021ab}:
    \begin{equation}\label{eq:d}
        d = \left( -0.018 \, g \bar{g}_0 + 0.009 \, g \bar{g}_1 - 0.023 \, g \bar{g}_2 \right)~e~\mathrm{fm}.
    \end{equation}
    In addition, we have computed the EDM of $^{19}$F also with the NN-N$^{4}$LO + 3N$^*_\mathrm{lnl}$ [see Fig.\ 5(b) in Supplemental Material~\cite{SM}] and obtained results consistent with Equation~\eqref{eq:d} well within the 30\% uncertainty quoted in Ref.~\cite{froese2021ab}.

    At an applied electric field of 58~V/cm \cite{Roussy2023an}, Equations~\eqref{eq:hfEDM} and \eqref{eq:d} translate to:
    \begin{equation}\label{eq:hfEDM_number}
        h f_\mathrm{EDM} = \left( 0.53 \, g \bar{g}_0 - 0.27 \, g \bar{g}_1 + 0.68 \, g \bar{g}_2 \right) \, h~\mathrm{Hz}.
    \end{equation}
    The coefficients in Equation~\eqref{eq:hfEDM_number} are much smaller than those in Equation~\eqref{eq:S19F_number}. Hence, we shall neglect the effect of the nuclear EDM of $^{19}$F in subsequent discussions. 
    
    It is important to note a specific detail of the HfF$^+$ experiment. The frequency reported in Ref.~\cite{Roussy2023an} corresponds to a particular experimental configuration optimized for sensitivity to the electron’s EDM and the NSM of $^{19}$F. This configuration exploits the fact that both observables depend on the projection of the electron and nuclear spins onto the internuclear axis. In contrast, the observable associated with the nuclear EDM of $^{19}$F depends on the projection of the nuclear spin onto the direction of the applied external electric field. As a result, the frequency reported in Ref.~\cite{Roussy2023an} is not sensitive to the nuclear EDM of $^{19}$F. In the terminology of Refs.~\cite{Roussy2023an,caldwell2023systematic}, the nuclear EDM contributes to the $f^B$ channel, whereas the reported frequency pertains to the $f^{DB}$ channel, which is sensitive to the electron's EDM and NSM of $^{19}$F. For a detailed discussion of these channels and their respective sensitivities, the reader is referred to Refs.~\cite{Roussy2023an,caldwell2023systematic}.
    
\vspace{2mm}    
\PRLsection{Discussion}
    Assuming that both the electron's EDM and the scalar–pseudoscalar nucleon–electron coupling are zero, we can place a bound on the NSM of $^{19}$F: 
    \begin{equation}\label{eq:S19F_bound}
        |S(^{19}\mathrm{F})| < 9.0 \times 10^{-9}~e~\mathrm{fm}^3 \quad \text{(90\% confidence level)}. 
    \end{equation}
    The conversion to a 90\% confidence level follows the methodology described in Ref.~\cite{Roussy2023an}. 
    
    If we further assume that each $\mathcal{P,T}$-violating observable in Equation~\eqref{eq:S19F} individually accounts for the entire NSM, we can derive bounds on these observables, as listed in Table~\ref{tab:TVobservables}.
    \begin{table}[tb]
        \caption{\textbf{Upper bounds (90\% confidence level) on $\mathcal{P,T}$-violating observables.} Definitions of $\bar{g}_{0,1,2}$ are provided in Equation~\eqref{eq:S19F}. The bounds are derived using the $|S(^{19}\mathrm{F})|$ limit from Equation~\eqref{eq:S19F_bound} and the coefficients in Equation~\eqref{eq:S19F}. Each bound is calculated under the assumption that the corresponding observable is the sole contributor to the nuclear Schiff moment of $^{19}$F. For the conversions, we use $g \approx 13.5$ \cite{ENGEL201321}. These bounds are consistent with previous constraints~\cite{Graner2016reduced}. }
        \label{tab:TVobservables}
        \centering
        \begingroup
        \setlength{\tabcolsep}{4pt}
        \centering
        \rowcolors{2}{gray!20}{white}
        \begin{tabular}{c l}
            \toprule
            Quantity & Limit \\
            \midrule
            $|\bar{g}_0|$ & $2.3 \times 10^{-8}$ \\
            $|\bar{g}_1|$ & $2.8 \times 10^{-8}$ \\
            $|\bar{g}_2|$ & $3.3 \times 10^{-8}$ \\
            \bottomrule
        \end{tabular}
        \endgroup             
    \end{table}
    
    Using $\bar{g}_0 \approx -17.2 \times 10^{-3} \, \bar{\theta}$ \cite{mulder2025probing} and $\bar{g}_1 \approx 3.4 \times 10^{-3} \, \bar{\theta}$ \cite{bsaisou2015nuclear, mulder2025probing}, while neglecting the highly suppressed coupling $\bar{g}_2$ \cite{de_Vries_2015}, we can also place a bound on the QCD $\theta$ term related to the strong $\mathcal{CP}$ problem \cite{POSPELOV1999243, peccei1989strong} :
    \begin{equation}
        |\bar{\theta}| < 1.6 \times 10^{-6} \quad \text{(90\% confidence level)}. 
    \end{equation}
    It is worth noting that the constraint on $\bar{\theta}$ derived from the limit on the scalar–pseudoscalar nucleon–electron coupling measured in the HfF$^+$ experiment is more stringent than the bound obtained from the nuclear Schiff moment (NSM) of $^{19}$F \cite{mulder2025probing}.

    Many of the molecules listed in Table~\ref{tab:WS} are under investigation for measurements of the NSM, e.g., TlF \cite{Grasdijk_2021}, RaF \cite{udrescu2024precision}, and $^{227}$ThF$^+$ (TRIUMF). These systems are chosen primarily for their strong sensitivity to NSMs associated with heavy nuclei \cite{chen2024relativistic}, where relativistic effects and nuclear structure lead to an intrinsic enhancement of the NSM signal. As a result, the sensitivity of these molecules to the NSMs of heavy elements can be up to four orders of magnitude greater than to that of $^{19}$F. Therefore, the contribution of the NSM of $^{19}$F to the interpretation of EDM measurements in such systems is expected to be at the level of $10^{-4}$. Although the $^{19}$F NSM exhibits reduced sensitivity to underlying $\mathcal{P,T}$-violating interactions compared to these heavier nuclei, its sensitivity coefficients are more accurately determined, as enabled by the nuclear \textit{ab initio} calculation in the present work, and can be used with greater confidence in constraining fundamental symmetry-violating physics. 
    
    In the future, and in line with ongoing efforts to extend the frontier of \textit{ab initio} nuclear theory to the heaviest nuclear systems, \textit{ab initio} approaches will aim to calculate NSMs of heavy, octupole-deformed nuclei as well. This remains a challenging task that, to date, has been addressed only using phenomenological nuclear models. Recent advances in the treatment of 3N forces~\cite{miyagi2022converged}, global ground-state calculations~\cite{stroberg2021ab}, and applications in the $^{208}$Pb region~\cite{hu2022ab} have pushed \textit{ab initio} methods toward heavier mass ranges. Once remaining many-body challenges --- such as effectively summing over all opposite-parity intermediate states coupled by the Schiff operator or treating nuclear deformation in the mass region beyond the doubly magic nucleus $^{208}$Pb --- are overcome, the NSMs of heavy, octupole-deformed nuclei are expected to come within reach of nuclear \textit{ab initio} frameworks. Promising developments in this direction include, for example, the use of non-spherical references to capture deformation effects~\cite{yao2020ab,hu2024ab}. In this context, our (quasi-)exact calculation of the NSM in $^{19}$F will provide a critical benchmark for validating and guiding future \textit{ab initio} efforts —-- particularly those that involve truncations in the model space, Hamiltonian, or operators —-- when targeting symmetry-violating phenomena such as the NSM in heavier mass regions of the nuclear chart that are beyond the reach of the NCSM.

    Moreover, in molecules where $^{19}$F is the only nucleus with non-zero spin, the nuclear-spin dependent $\mathcal{P,T}$-violating effects in the molecular EDM arise predominantly from the NSM of $^{19}$F. Proposed experiments using such molecules, e.g., Refs.~\cite{Fitch2021methods, vutha2018orientation}, aim to leverage long trapping times and high molecular densities, with projected frequency sensitivities at the $10^{-8}~\mathrm{Hz}$ level, far exceeding current experimental precision. Such advances could yield competitive or complementary bounds on $\mathcal{P,T}$-violating interactions, rivaling the most stringent existing constraints \cite{Graner2016reduced, nedm2020}.

\PRLsection{Conclusion}
    We have presented the first \textit{ab initio} calculations of a nuclear Schiff moment (NSM), focusing on $^{19}$F. By integrating (i) our nuclear structure calculations of the $^{19}$F NSM, (ii) quantum chemistry analyses of molecular sensitivity to the NSM, and (iii) a detailed interpretation of the recent measurement of the electric dipole moment in HfF$^+$ \cite{Roussy2023an}, we have established the first bound on the NSM of $^{19}$F. From this bound, we derive corresponding limits on the pion–nucleon–nucleon coupling which, while consistent with existing more stringent constraints, are obtained here for the first time within a nuclear \textit{ab initio} framework. This work demonstrates the power of \textit{ab initio} approaches in connecting fundamental symmetries with experimental observables and paves the way for interpreting future precision measurements aimed at uncovering new sources of violations of fundamental symmetry in the nuclear sector.

\PRLsection{Acknowledgment}
    This work is supported by the Natural Sciences and Engineering Council of Canada (NSERC) Grant No. SAPIN-2022-00019 and SAPPJ/42-2023, as well as the National Science Foundation under Grant No.\ PHY-2309253 (L.\ C.). TRIUMF receives federal funding via a contribution agreement with the National Research Council of Canada. K.~B.~Ng acknowledges support from the Banting Postdoctoral Fellowship BPF-198564. Computing support came from an INCITE Award on the Frontier supercomputer of the Oak Ridge Leadership Computing Facility (OLCF) at ORNL, from the LLNL LC, and from the Digital Research Alliance of Canada. We are grateful for insightful discussions with Jason Holt. 
    
\bibliography{biblio}

\end{document}


\title{Supplemental Material:\\Nuclear Schiff moment of the fluorine isotope $^{19}$F}

\author{Kia Boon Ng}
\email{kbng@triumf.ca}
\affiliation{TRIUMF, Vancouver, British Columbia, V6T 2A3, Canada}

\author{Stephan Foster}
\affiliation{TRIUMF, Vancouver, British Columbia, V6T 2A3, Canada}
\affiliation{Department of Engineering Physics, McMaster University, Hamilton, Ontario, L8S 4M1, Canada}

\author{Lan Cheng}
\affiliation{Department of Chemistry, The Johns Hopkins University, Baltimore, MD 21218, USA}

\author{Petr Navr\'atil}
\affiliation{TRIUMF, Vancouver, British Columbia, V6T 2A3, Canada}

\author{Stephan Malbrunot-Ettenauer}
\affiliation{TRIUMF, Vancouver, British Columbia, V6T 2A3, Canada}
\affiliation{Department of Physics, University of Toronto, Toronto, Ontario, M5S 1A7, Canada}

\maketitle

\section{Nuclear calculations}

We have performed nuclear \textit{ab initio} calculations of the $^{19}$F nuclear Schiff moment (NSM) using the no-core shell model (NCSM) approach \cite{BARRETT2013131} in combination with the Lanczos strength method \cite{Haydock1974, Marchisio2003}. As input, we used primarily two sets of parity-conserving (PC) $\chi$EFT nucleon-nucleon (NN) and three-nucleon (3N) interactions: NN-N$^4$LO~\cite{Entem2017} + 3N$^*_\mathrm{lnl}$~\cite{Kravvaris2023,Jokiniemi2024,Girlanda2011} (also denoted as NN N$^4$LO + 3N$_\mathrm{lnlE7}$ in some of the figures) and NN-N$^3$LO~\cite{Entem2003} + 3N$_\mathrm{lnl}$~\cite{Soma2020}. In the former, an additional sub-leading contact term ($E_7$) enhancing the spin-orbit strength~\cite{Girlanda2011} has been introduced to the 3N force. The $E_7$ low-energy constant has been adjusted to improve the description of the excitation energies of $^6$Li, in particular of the first excited state $3^+$, $T=0$. Both high-precision NN-N$^4$LO and NN-N$^3$LO interactions use a 500 MeV regulator cutoff. The interactions have been softened by the Similarity Renormalization Group (SRG) technique \cite{Bogner2007} with the SRG induced three-nucleon terms fully included. In the present study, we use the evolution parameter $\lambda_{\rm SRG}=1.8$ fm$^{-1}$ for the NN-N$^4$LO+3N$^*_{\rm lnl}$ interaction and $\lambda_{\rm SRG}=2.0$ fm$^{-1}$ for the NN-N$^3$LO+3N$_{\rm lnl}$. As shown below, we have checked that the observables are insensitive to the variation of the $\lambda_{\rm SRG}$ parameter between 1.8 and 2.0 fm$^{-1}$. To further test sensitivity of the NSM to the input PC chiral interactions, we also used lower-order NN-N$^3$LO$_{\rm new}$ 500~\cite{Entem2017} + 3N$_\mathrm{lnl}$~\cite{PhysRevC.102.024616,Kravvaris2023} and NN-N$^2$LO 500~\cite{Entem2017} + 3N$_\mathrm{lnl}$~\cite{PhysRevC.102.024616,Kravvaris2023} interactions.

The NCSM calculations for the $^{19}$F positive-parity states have been performed up to $N_{\rm max}{=}6$ basis space where the $m$-scheme dimension reach 1.35 billion, which is, to the best of our knowledge, the largest such NCSM calculation ever performed with non-truncated 3N interaction. For the negative parity states, we have been able to reach $N_{\rm max}{=}5$ space only ($m$-scheme dimension 189 million). The full $N_{\rm max}{=}7$ space with a dimension of 8.48 billion~\cite{Johnson2025} is beyond our technical capability. However, we have been able to perform limited calculations of the NSM using importance truncated NCSM (IT-NCSM)~\cite{Roth2007,Roth2009,Kruse2013} $N_{\rm max}{=}7$ space as described below.

\begin{figure}
\includegraphics[width=0.6\textwidth]{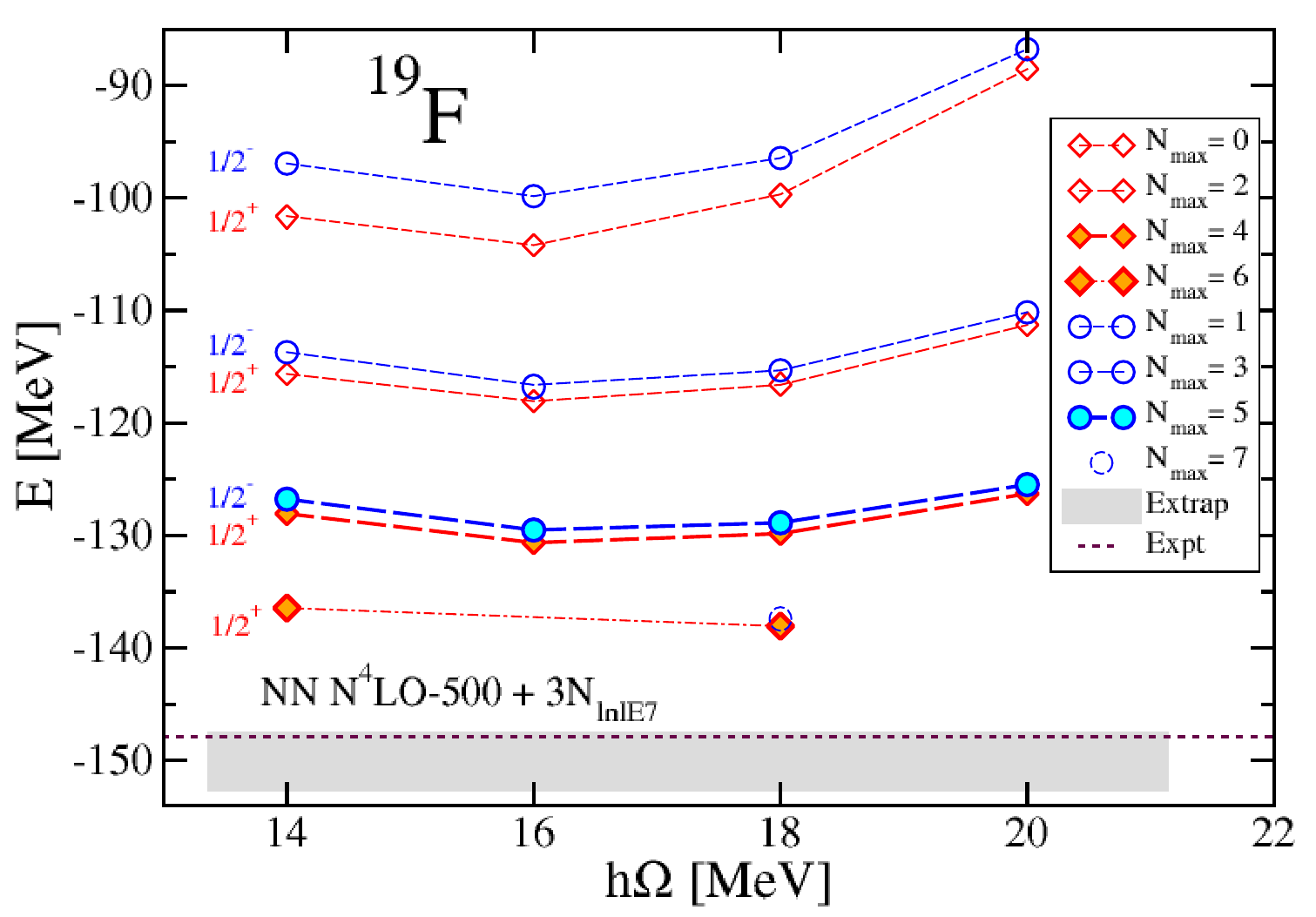}
\caption{\label{fig:F19-gs-energy} $^{19}$F $1/2^+$ ground-state and $1/2^-$ first excited state energy dependence on the HO frequency for different sizes of the NCSM model space. The SRG-evolved NN-N$^4$LO + 3N$^*_\mathrm{lnl}$ interaction was used. The dotted line corresponds to the $1/2^+$ ground-state experimental energy. See the text for further details.}
\end{figure}
In Fig.~\ref{fig:F19-gs-energy}, we present the $^{19}$F $1/2^+$ ground-state and $1/2^-$ first excited state energy dependence on the HO frequency for different sizes of the NCSM model space for calculations with the NN-N$^4$LO + 3N$^*_\mathrm{lnl}$ interaction. With increasing basis size, the energy decreases and the harmonic-oscillator (HO) frequency becomes flatter as expected. Also, the excitation energy of the $1/2^-$ state decreases with $N_{\rm max}$. The $1/2^+$ energy minimum is between $\hbar\Omega{=}16$ and 18 MeV, which is where we mostly focus our NSM calculations. The gray band corresponds to the exponentially extrapolated $1/2^+$ energy using $E(N_{\rm max})=a + b\,e^{-cN_{\rm max}}$ with its width representing the uncertainty obtained from extrapolations of the $N_{\rm max}{=}2,4,6$ points at $\hbar\Omega{=}14$ and 18 MeV. 

\begin{figure}
\includegraphics[width=0.6\textwidth]{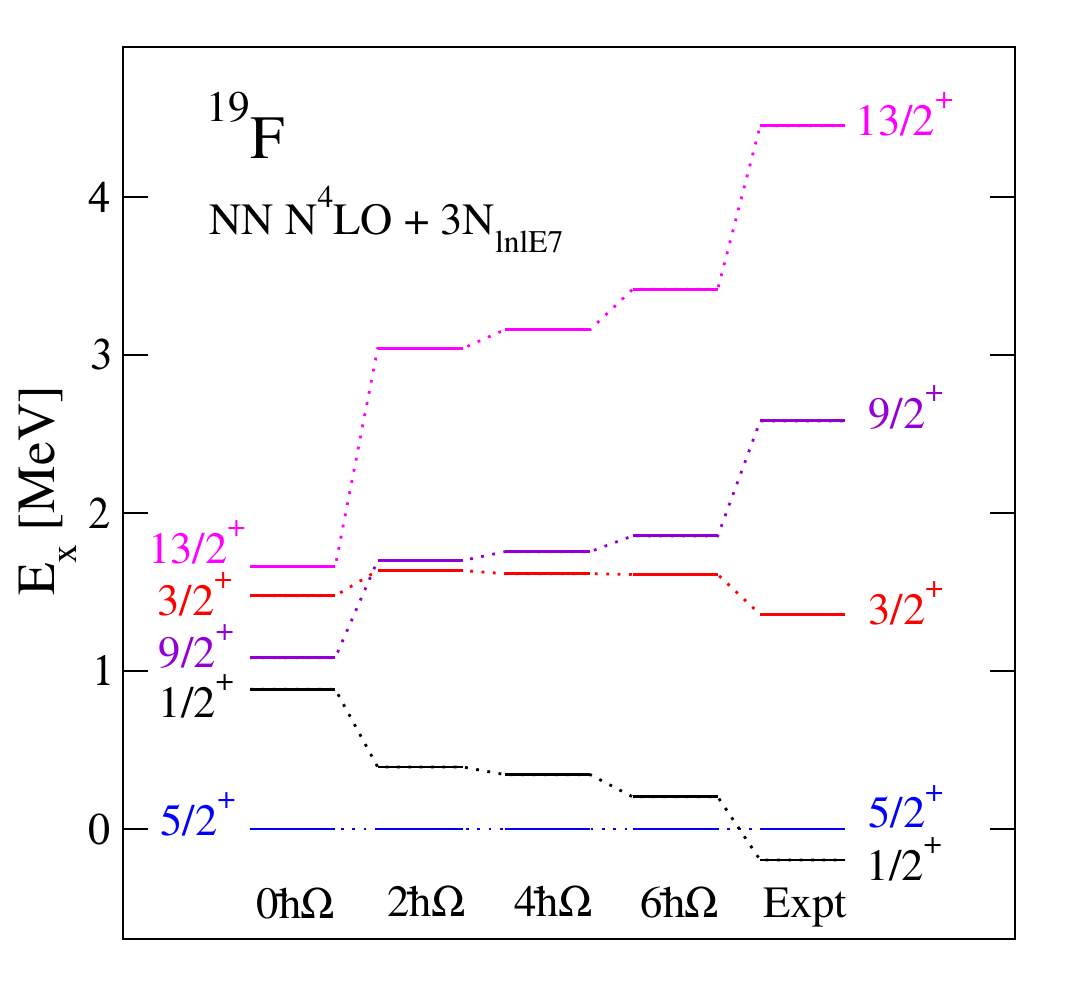}
\caption{\label{fig:F19-exc-energy} Excitation energy dependence on the NCSM model space size for the $^{19}$F low-lying positive parity states. The SRG-evolved NN-N$^4$LO + 3N$^*_\mathrm{lnl}$ interaction and the HO frequency of $\hbar\Omega{=}14$ MeV were used. See the text for further details.}
\end{figure}
The $^{19}$F low-lying positive-parity state excitation energy dependence on the basis size is shown in Fig.~\ref{fig:F19-exc-energy}. Up to $N_{\rm max}{=}6$, the calculated ground state is $5/2^+$, consequently, we present the excitation energies with respect to that state. With increasing basis size, all the energy levels improve their agreement with experiment. It is not straightforward to project, but it seems likely that the NN-N$^4$LO + 3N$^*_\mathrm{lnl}$ interaction will predict the correct $1/2^+$ ground state at $N_{\rm max}\rightarrow\infty$.

\begin{figure}
\includegraphics[width=0.6\textwidth]{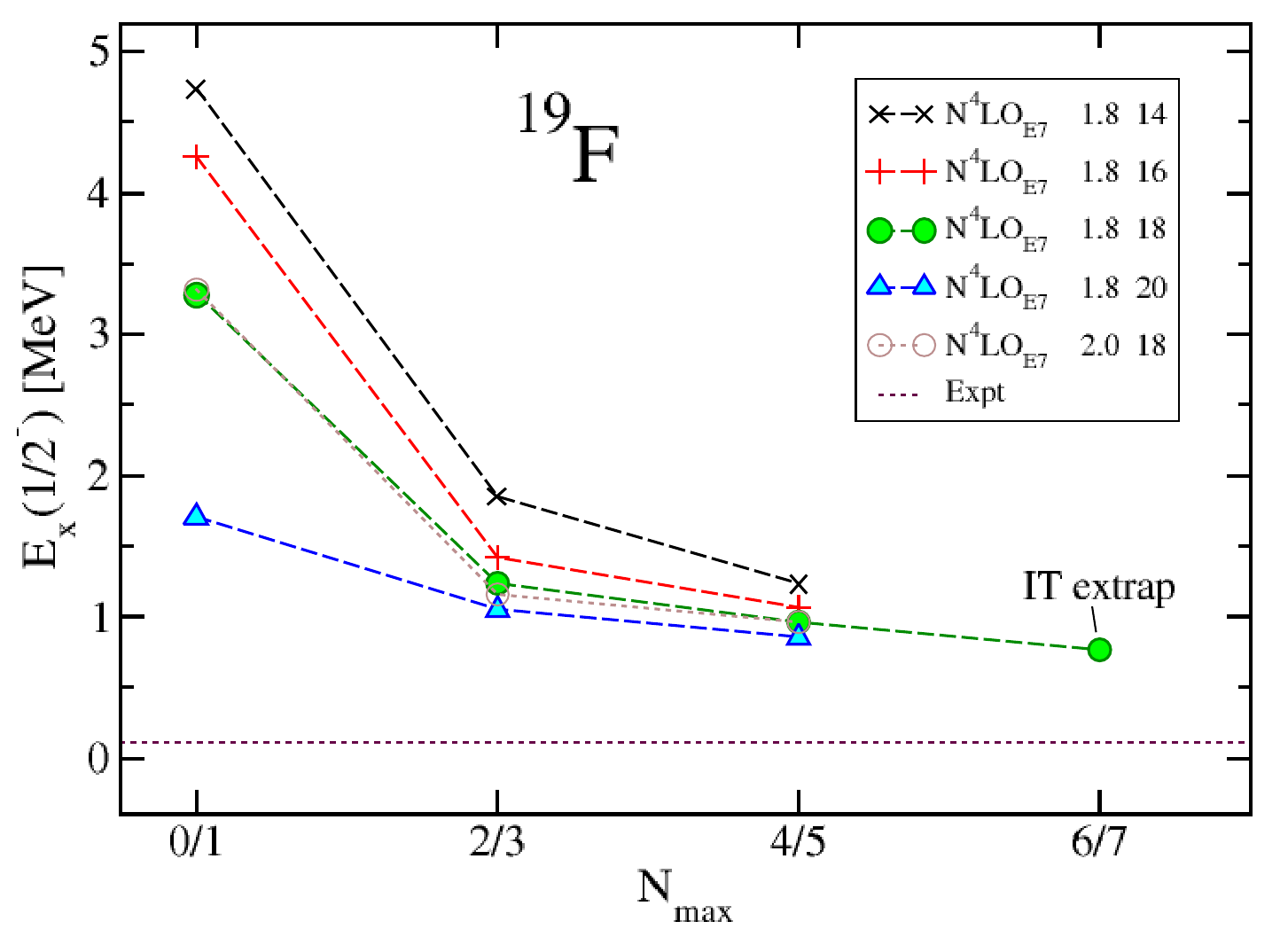}
\caption{\label{fig:F19-1m-exc-energy} The $^{19}$F $1/2^-_1$ state excitation energy dependence on the NCSM model space size for a range of HO frequencies shown in the last column of the legend (in MeV). The SRG-evolved NN-N$^4$LO + 3N$^*_\mathrm{lnl}$ interaction was used. The $\lambda_{\rm SRG}$ parameter is shown in the middle column of the legend (in fm$^{-1}$). See the text for further details.}
\end{figure}
\begin{figure}[hbt!]
\begin{subfigure}{.475\linewidth}
  \includegraphics[width=\linewidth]{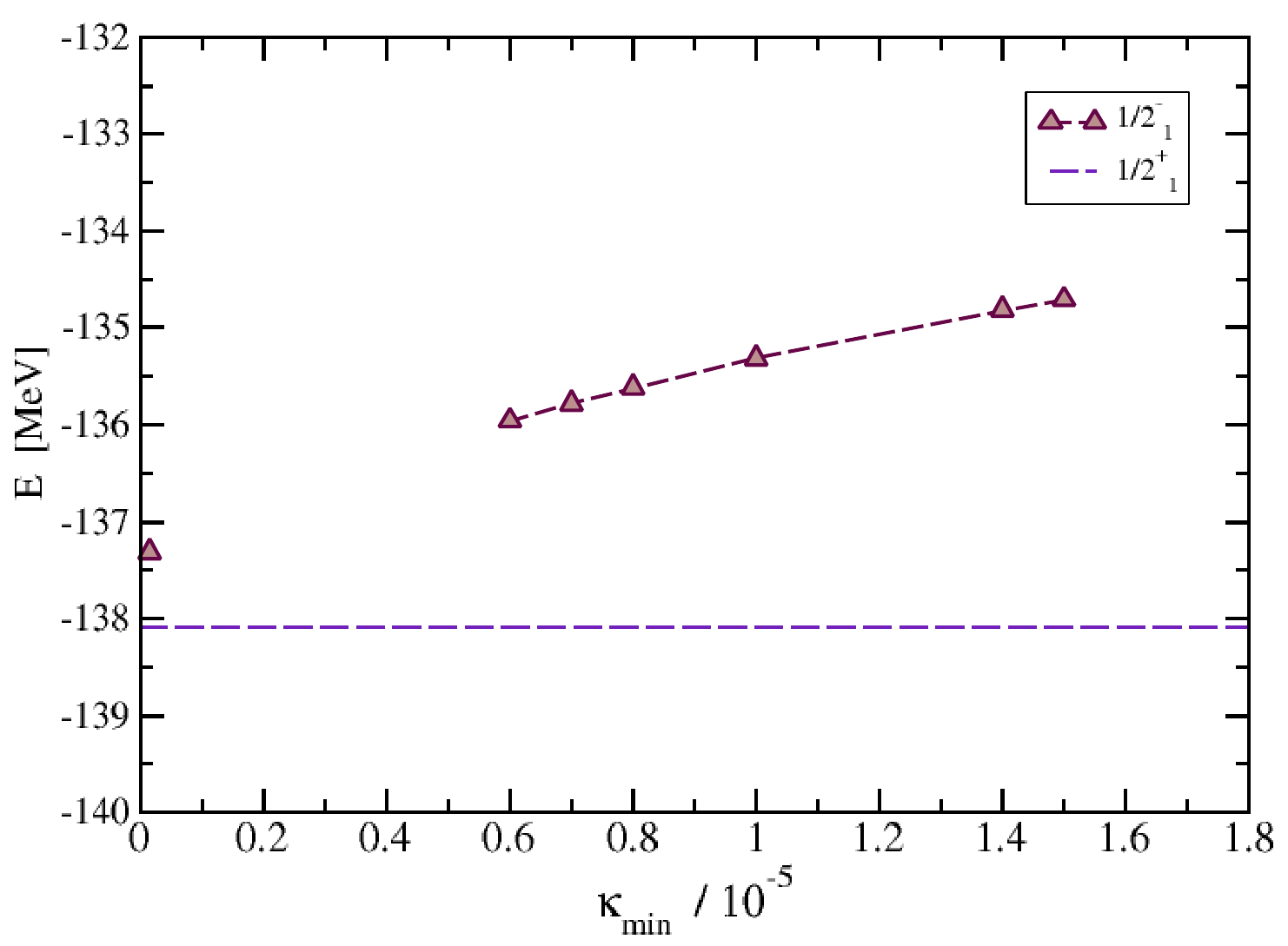}
  \caption{}
  \label{1}
\end{subfigure}\hfill 
\begin{subfigure}{.475\linewidth}
  \includegraphics[width=\linewidth]{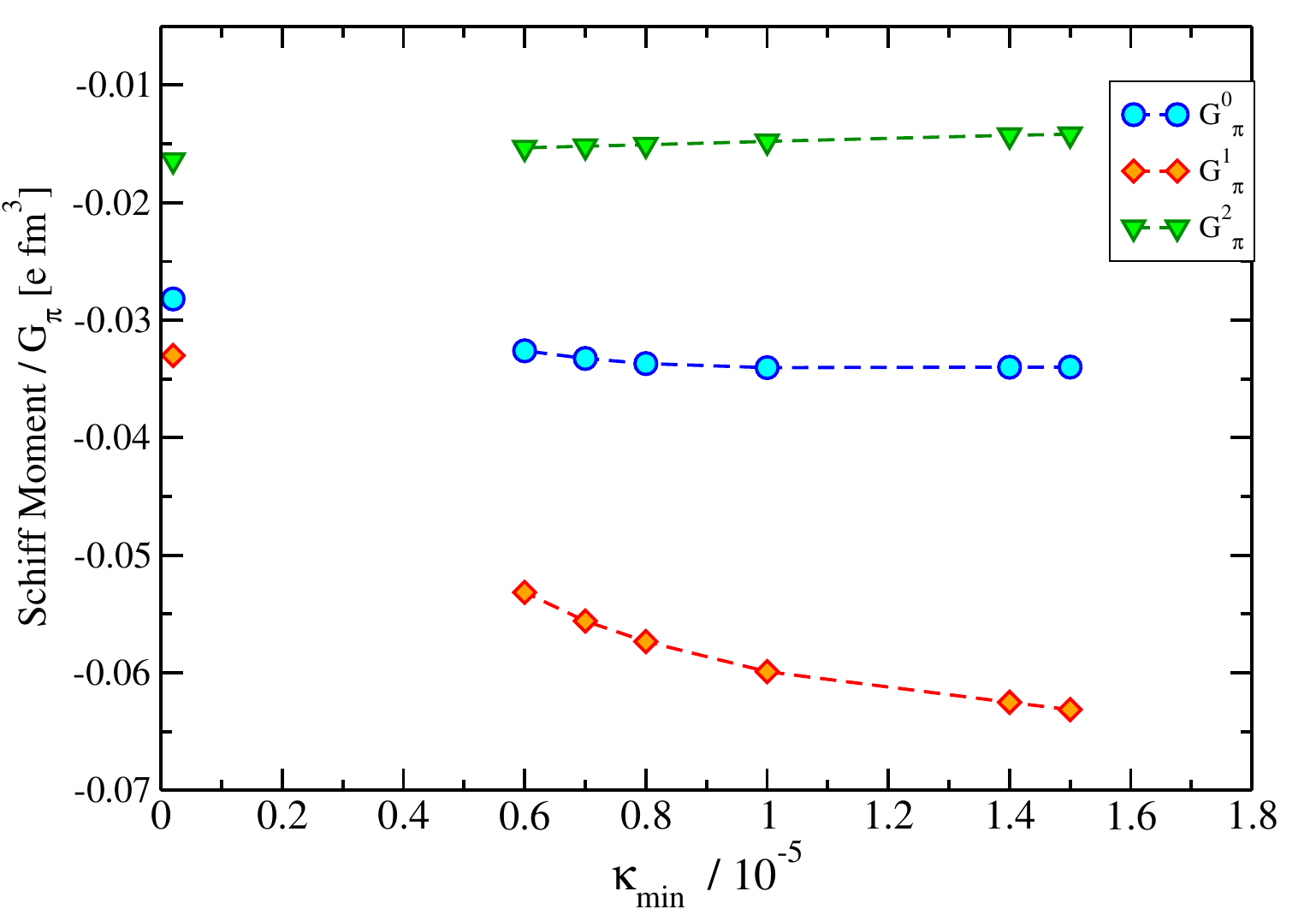}
  \caption{}
  \label{2}
\end{subfigure}
\caption{Dependence of the $^{19}$F $1/2^-_1$ state energy (a) and the NSM one-$\pi$-exchange contributions due to the $1/2^-_1$ state (b) on the IT-NCSM basis-state importance measure $\kappa_{\rm min}$. The SRG-evolved NN-N$^4$LO + 3N$^*_\mathrm{lnl}$ interaction in $N_{\rm max}{=}7$ space and the HO frequency of $\hbar\Omega{=}18$ MeV were used. The polynomial extrapolated $\kappa_{\rm min}\rightarrow0$ points are also shown. The dashed line in the panel (a) denotes the corresponding full-space $N_{\rm max}{=}6$ $1/2^+_1$ energy. See the text for further details.}
\label{fig:kappa_min_extrap}
\end{figure}

The $^{19}$F $1/2^-_1$ state excitation energy dependence on the NCSM model space size is presented in Fig.~\ref{fig:F19-1m-exc-energy}. The energy is computed by matching the $1/2^+_1$ state obtained at $N_{\rm max}$ with the $1/2^-_1$ state at $N_{\rm max}{+}1$. The $N_{\rm max}{=}7$ energy, also shown in Fig.~\ref{fig:F19-gs-energy}, was obtained by extrapolating the IT-NCSM basis-state importance measure $\kappa_{\rm min}$ to zero (see Fig.~\ref{fig:kappa_min_extrap} (a)), while a full-space calculation was performed for the corresponding $N_{\rm max}{=}6$ $1/2^+$ energy. We find a decreasing excitation energy with $N_{\rm max}$ and a decreasing $\Omega$ dependence as expected. The experimental value of 0.1099 MeV is shown by the dotted line. It is challenging to extrapolate the converged excitation energy given the limited number of points. As we show later, the $1/2^-_1$ state has an $\alpha$ cluster structure suggesting a slow convergence with $N_{\rm max}$. In Fig.~\ref{fig:F19-1m-exc-energy}, a negligible dependence of the $1/2^-_1$ state excitation energy on the $\lambda_{\rm SRG}$ parameter is demonstrated in the $\hbar\Omega{=}18$ MeV calculations.

We note that we have computed excitation energies of other negative parity states and they match experimental situation closely. For example, at $N_{\rm max}{=}5, \hbar\Omega{=}14$ MeV we obtained $E_{\rm x}(5/2^-_1){=}1.01$ MeV with respect to (wrt) the $1/2^-_1$ (expt. 1.24 MeV), $E_{\rm x}(3/2^-_1){=}1.42$ MeV wrt the $1/2^-_1$ (expt. 1.35 MeV), $E_{\rm x}(9/2^-_1){=}3.35$ MeV wrt the $1/2^-_1$ (expt. 3.92 MeV), $E_{\rm x}(7/2^-_1){=}3.84$ MeV wrt the $1/2^-_1$ (expt. 3.89 MeV). The $1/2^-_2$ state appears in experiment at 6.43 MeV wrt the $1/2^-_1$. Our $N_{\rm max}{=}5, \hbar\Omega{=}18$ MeV calculation place it at $6.37$ MeV.

Overall, we find a quite reasonable agreement of calculated $^{19}$F energies with experiment. The present results confirm a good performance of the chiral NN-N$^4$LO + 3N$^*_\mathrm{lnl}$ interaction in light systems as observed in earlier NCSM and NCSM with continuum (NCSMC) calculations~\cite{Kravvaris2023,Jokiniemi2024}. In fact, it is the best performing interaction available to us currently.

The leading-order (LO) $\mathcal{P,T}$-violating (PTV) NN interaction due to the one-pion exchange is taken in the form introduced in Ref.~\cite{Liu2004}, which is equivalent to the LO $\chi$EFT $\mathcal{P,T}$-violating NN interaction~\cite{deVries2020}. Unnatural parity states are admixed into the ground state by this PTV NN interaction
\begin{equation}\label{gswf}
  |A, \, {\rm g.s.},\; I \rangle = |A, \, {\rm g.s.},\; I^\pi \rangle + \sum_\lambda |A,\, \lambda, \; I^{-\pi}\rangle 
  \times \frac{\langle A,\, \lambda, \; I^{-\pi}| V_{\rm NN}^{\rm PTV}|A, \, {\rm g.s.}, \; I^\pi \rangle}{E_{\rm g.s.}^{I^\pi}-E_\lambda^{I^{-\pi}}}  \; ,
\end{equation}
which incorporates the symmetry breaking necessary to generate a Schiff Moment. The NSM is evaluated according to Eq.~(3) in the main text with the operator ${\bf S}$ given in Eq.~(2) in the main text and the wave function from Eq.~(\ref{gswf}). To avoid the sum over intermediate states in Eq.~(\ref{gswf}), we rather invert Eq.~(4) in the main text by applying the Lanczos strength method. Note that we shift the calculated $E_{\rm g.s.}^{I^\pi}$ energy to match experimental excitation energy of the lowest opposite-parity state (here $1/2^-_1$). The size of this shift can be inferred from Fig.~\ref{fig:F19-1m-exc-energy}. The fact that the used chiral NN+3N interactions overestimate the splitting of the $1/2^+_1$ and $1/2^-_1$ states by a few hundreds of keV does not limit the confidence in the correctness of the matrix elements of the Schiff operator ${\bf S}$ and the PTV NN interaction. This is because the relevant scale for the NCSM wave functions is given by the total energies, which are of the order of 140 MeV (Fig.~\ref{fig:F19-gs-energy}). 

\begin{figure}[hbt!]
\begin{subfigure}{.475\linewidth}
  \includegraphics[width=\linewidth]{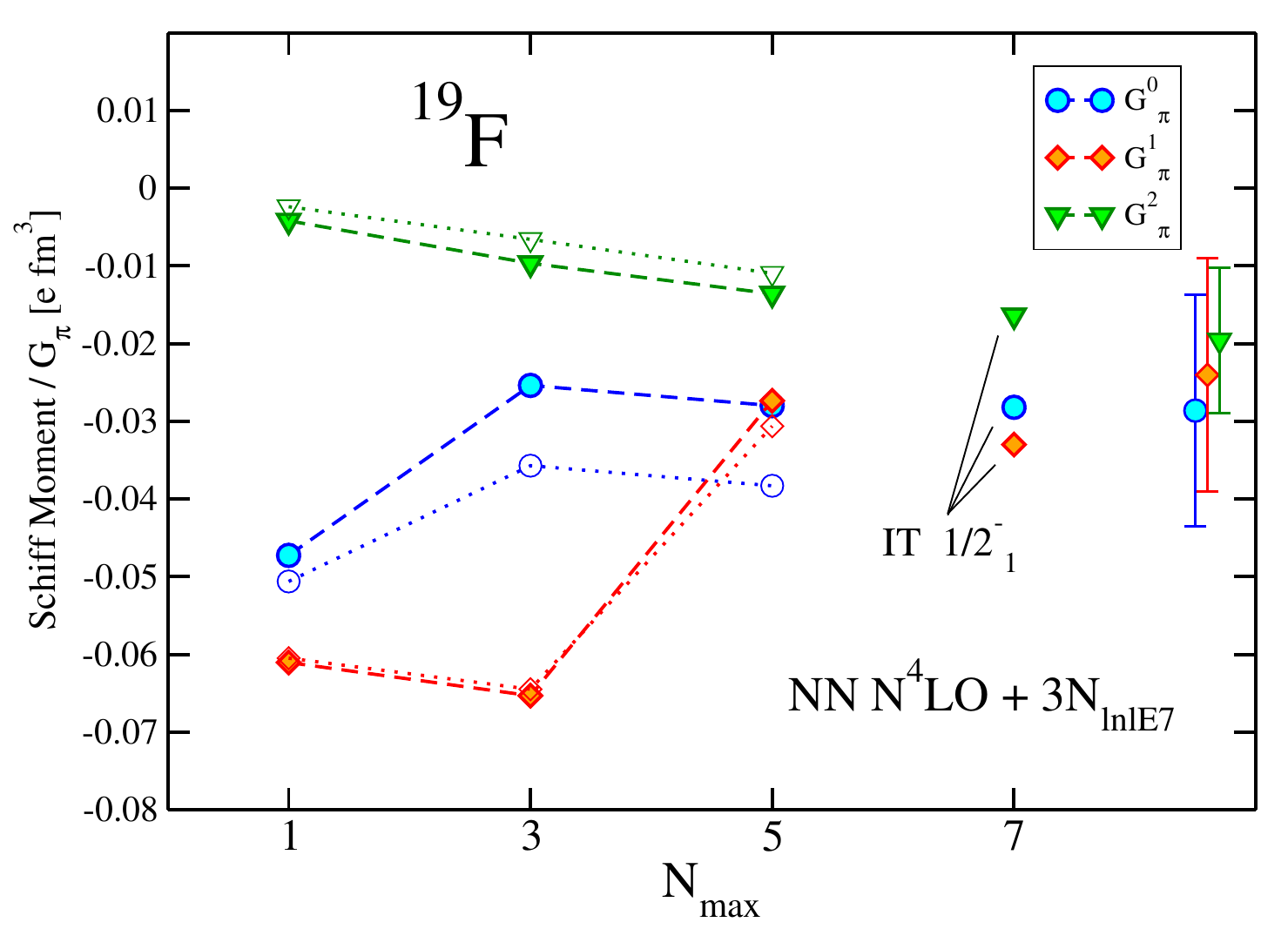}
  \caption{}
  \label{1}
\end{subfigure}\hfill 
\begin{subfigure}{.475\linewidth}
  \includegraphics[width=\linewidth]{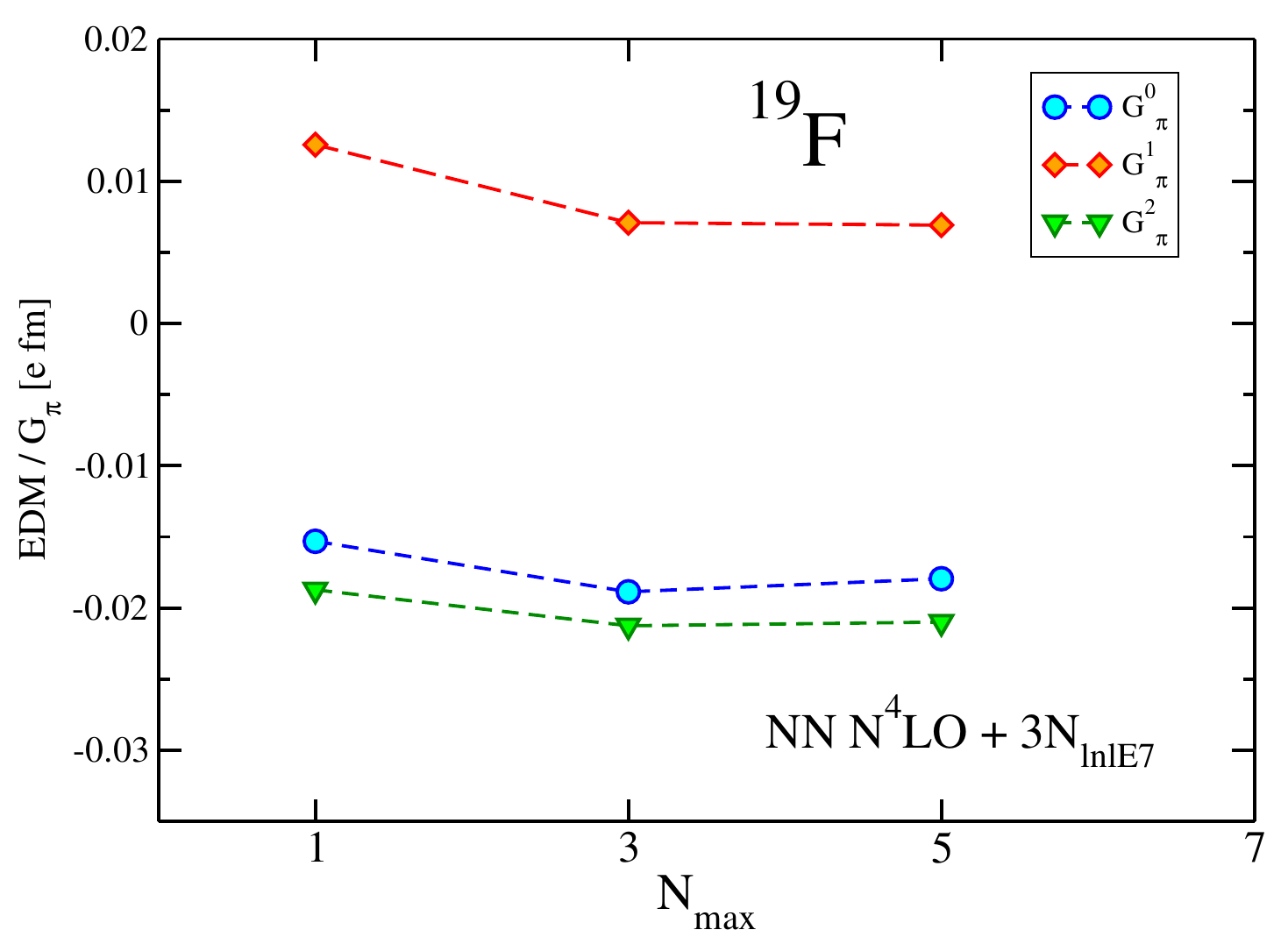}
  \caption{}
  \label{2}
\end{subfigure}
\caption{$^{19}$F NSM (a) and EDM (b) one-$\pi$-exchange contribution dependence on the model space size. The SRG-evolved NN-N$^4$LO + 3N$^*_\mathrm{lnl}$ interaction was used with HO frequencies of $\hbar\Omega{=}16$ MeV (dashed lines in panel (a)) and $\hbar\Omega{=}18$ MeV (panel (b) and the dotted lines and $N_{\rm max}{=}7$ results in panel (a)). The $N_{\rm max}{=}7$ NSM results were obtained within the extrapolated IT-NCSM with the $1/2^-_1$ state only (Fig.~\ref{fig:kappa_min_extrap} (b)). The renormalized $N_{\rm max}{=}7$ NSM values, which we consider our final results, are shown with their theoretical uncertainties on the right of panel (a). Details are given in the text.}
\label{fig:F19-SNM_Nmax-dep}
\end{figure}
\begin{figure}[hbt!]
\begin{subfigure}{.475\linewidth}
  \includegraphics[width=\linewidth]{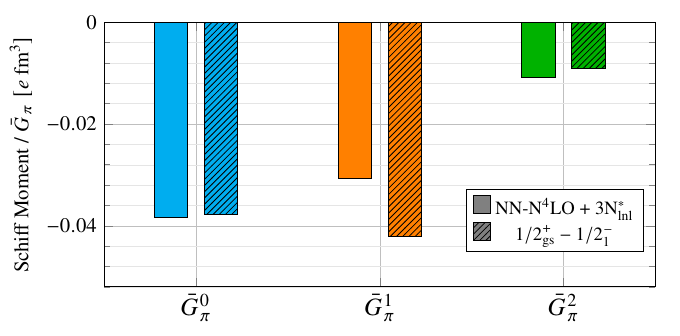}
  \caption{}
  \label{1}
\end{subfigure}\hfill 
\begin{subfigure}{.475\linewidth}
  \includegraphics[width=\linewidth]{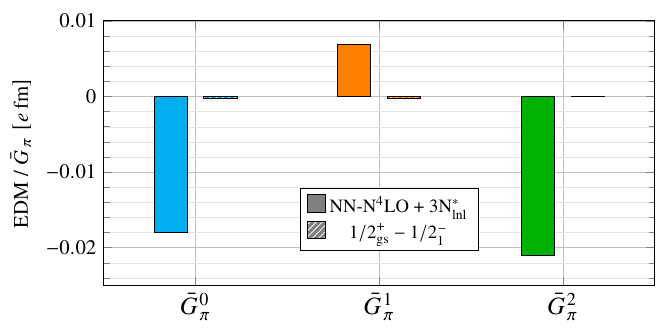}
  \caption{}
  \label{2}
\end{subfigure}
\caption{$^{19}$F NSM (a) and EDM (b) one-$\pi$-exchange contributions. Calculations including all $1/2^-$ states are compared to those with the $1/2^-_1$ state only. The SRG-evolved NN-N$^4$LO + 3N$^*_\mathrm{lnl}$ interaction in $N_{\rm max}{=}5$ space and the HO frequency of $\hbar\Omega{=}18$ MeV were used. See the text for further details.}
\label{fig:PV_moments_full-vs-1m-contr}
\end{figure}
\begin{figure}
\includegraphics[width=0.9\textwidth]{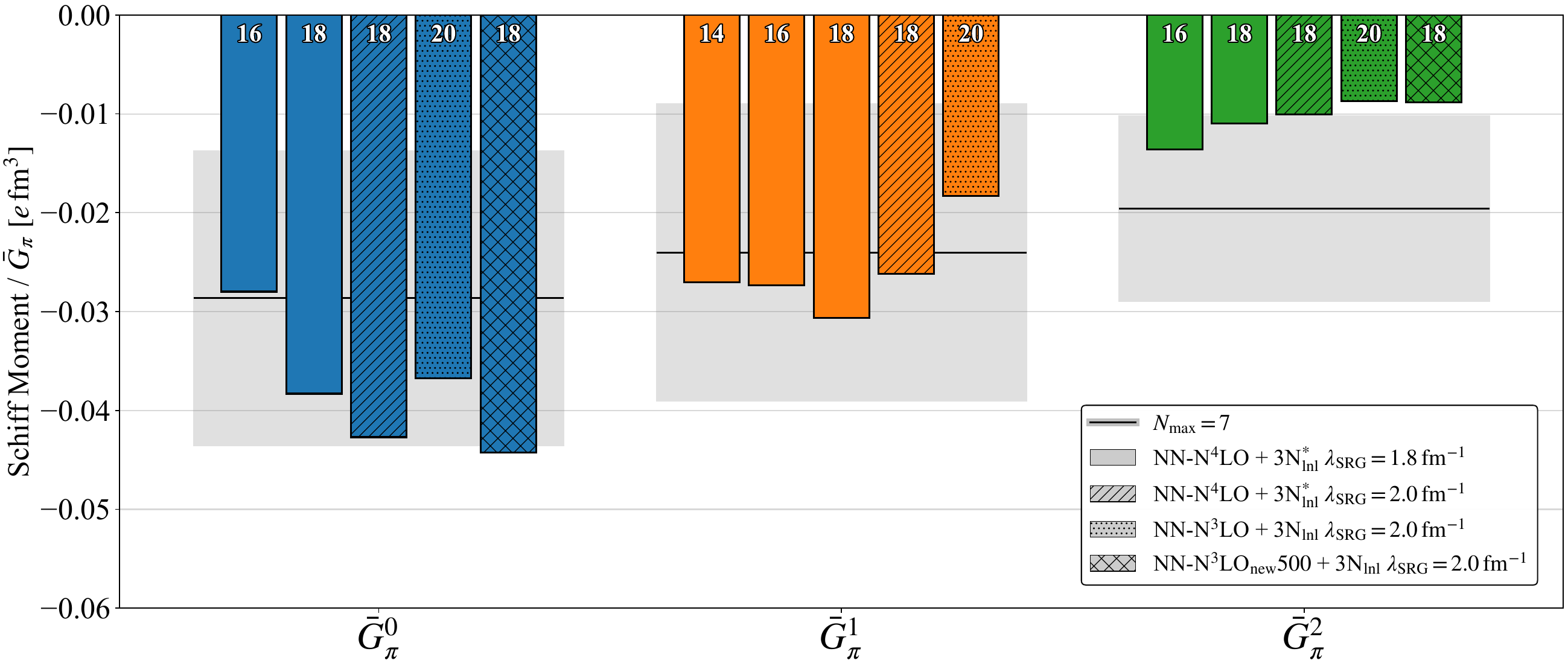}
\caption{\label{fig:F19-SNM_interaction-dep} Calculated $^{19}$F NSM one-$\pi$-exchange contributions. The bars represent the $N_{\rm max}{=}5$ results for various HO frequencies, SRG parameters and interactions as shown in the legend. The $N_{\rm max}{=}7$ results with theoretical uncertainties are shown by solid lines and gray bands, respectively. Details are given in the text.}
\end{figure}

Our $^{19}$F NSM results are presented in Figs.~\ref{fig:kappa_min_extrap} (b), \ref{fig:F19-SNM_Nmax-dep} (a), \ref{fig:PV_moments_full-vs-1m-contr} (a), and \ref{fig:F19-SNM_interaction-dep}. In addition, we show the $^{19}$F nuclear EDM results obtained with the NN-N$^4$LO + 3N$^*_\mathrm{lnl}$ interaction in Figs.~\ref{fig:F19-SNM_Nmax-dep} (b) and \ref{fig:PV_moments_full-vs-1m-contr} (b). Our nuclear EDM calculations show a very weak dependence on the basis size in line with observations reported in Ref.~\cite{froese2021ab}, and the $N_{\rm max}{=}5$ results agree closely with those of Ref.~\cite{froese2021ab} obtained with the NN-N$^3$LO + 3N$_{\rm lnl}$ interaction as pointed out in the main text. In contrast, the NSM calculations show a challenging $N_{\rm max}$ dependence for the isovector one-pion exchange contribution as seen in Fig.~\ref{fig:F19-SNM_Nmax-dep} (a). The calculations up to $N_{\rm max}{=}5$ (with matched $1/2^+_1$ states from $N_{\rm max}{-}1$ spaces) have been performed within the NCSM with the Lanczos strength method taking into account contributions of all $1/2^-$ excited states of $^{19}$F from a given basis space. As explained earlier, we have been able to reach the $N_{\rm max}{=}7$ space within the IT-NCSM. For technical reasons, the corresponding NSM calculations have been limited to the contribution from the lowest $1/2^-_1$ state only (labeled IT $1/2^-_1$ in Fig.~\ref{fig:F19-SNM_Nmax-dep} (a)), i.e., restricting the sum in Eq.~(\ref{gswf}) to $\lambda{=}1$. The matching $1/2^+_1$ eigenstate have been obtained in the full $N_{\rm max}{=}6$ space. Fig.~\ref{fig:kappa_min_extrap} (b) shows the NSM dependence on the importance measure $\kappa_{\rm min}$. The lowest $\kappa_{\rm min}$ calculations that we were able to reach, $0.6 \times 10^{-5}$, included 812 million basis states (i.e., full $N_{\rm max}{=}5$ space plus 623 million $N{=}7$ states) while the $\kappa_{\rm min}{=}0.7 \times 10^{-5}$ calculations included 702 million states. To avoid impact of the denominator energy shift, the polynomial extrapolations to $\kappa_{\rm min}\rightarrow 0$ have been done separately for the $1/2^+_1{\rightarrow}1/2^-_1$ matrix elements of the Schiff operator and the $\mathcal{P,T}$-violating NN interaction with the combined NSM contribution results shown on the left of Fig.~\ref{fig:kappa_min_extrap} (b).
It turns out that the $^{19}$F NSF is dominated by the contribution of the low-lying $1/2^-_1$ parity partner of the $1/2^+$ ground state as we have verified in the $N_{\rm max}{=}5$ and the smaller spaces. This is depicted in Fig.~\ref{fig:PV_moments_full-vs-1m-contr} (a). This is in contrast to the $^{19}$F nuclear EDM where the contribution of the $1/2^-_1$ state is negligible (Fig.~\ref{fig:PV_moments_full-vs-1m-contr} (b)). Taking advantage of this, we use the ratio of the full calculation and the $1/2^-_1$ contribution calculated in $N_{\rm max}{=}5$ space (Fig~\ref{fig:PV_moments_full-vs-1m-contr} (a)) to renormalize the $N_{\rm max}{=}7$ results and estimate the corresponding full space NSM. That estimate is shown on the right of Fig.~\ref{fig:F19-SNM_Nmax-dep} (a). It can be seen that the renormalized $N_{\rm max}{=}7$ values are consistent with the $N_{\rm max}{=}5$ ones for all three one-pion-exchange contributions. That gives us confidence to use these renormalized $N_{\rm max}{=}7$ values as our best estimates of the $^{19}$F NSM.

To estimate the theoretical uncertainties, we consider the variation of our results with the basis size, the HO frequency, the SRG evolution parameter, and the chiral interaction. The values used are shown in Fig.~\ref{fig:F19-SNM_interaction-dep} together with additional results. In particular, we take the differences of the $N_{\rm max}{=}5$ and the renormalized $N_{\rm max}{=}7$ results obtained at $\hbar\Omega{=}18$~MeV, the differences of the $\hbar\Omega{=}16$ MeV and $\hbar\Omega{=}18$ MeV results obtained at $N_{\rm max}{=}5$, the differences of results obtained with the $\lambda_{\rm SRG}{=}1.8$ fm$^{-1}$ and $\lambda_{\rm SRG}{=}2.0$ fm$^{-1}$ at $\hbar\Omega{=}18$ MeV, and the differences of results obtained with the NN-N$^4$LO+3N$^*_{\rm lnl}$ at $\hbar\Omega{=}18$ MeV and NN-N$^3$LO+3N$_{\rm lnl}$ interactions. Treating the above as independent, we arrive at the $^{19}$F NSM
\begin{equation}\label{eq:S19F_uncert} 
            S(^{19}\mathrm{F}) = (-2.9(15)\, g\bar{g}_0 - 2.4(15)\, g\bar{g}_1 - 2.0(9)\, g\bar{g}_2) \times 10^{-2}~e~\mathrm{fm}^3,
\end{equation}
where $g$ denotes the $\mathcal{P,T}$-conserving pion-nucleon-nucleon ($\pi$NN) coupling constant; and $\bar{g}_0$, $\bar{g}_1$, and $\bar{g}_2$ are the isoscalar, isovector, and isotensor $\mathcal{P,T}$-violating $\pi$NN coupling constants, respectively. In the figures, we use an abbreviated notation $\bar{G}^0_\pi{=}g\bar{g}_0$, $\bar{G}^1_\pi{=}g\bar{g}_1$, and $\bar{G}^2_\pi{=}g\bar{g}_2$. The results given in Eq.~(\ref{eq:S19F_uncert}) are also shown in Figs.~\ref{fig:F19-SNM_Nmax-dep} (a) and \ref{fig:F19-SNM_interaction-dep}. In the latter figure, we also present results obtained with the NN-N$^3$LO$_{\rm new}$ 500 + 3N$_\mathrm{lnl}$ interaction for the isoscalar and isotensor one-pion-exchange contributions that show consistency with the other values. We obtained similar results with the NN-N$^2$LO 500 + 3N$_\mathrm{lnl}$ interaction. However, these interactions suggest positive isovector one-pion-exchange NSM in the $N_{\rm msx}{=}5$ space (not shown in the figure). We find that the absolute value decreases significantly between the N$^2$LO and the N$^3$LO interaction calculations. Since these interactions have not been comprehensively tested in many-body calculations contrary to the two precision interactions we employed (NN-N$^4$LO+3N$^*_{\rm lnl}$ and NN-N$^3$LO+3N$_{\rm lnl}$), we use these results here only to highlight the increased sensitivity of the isovector one-pion-exchange NSM term to the input. 

\begin{figure}[hbt!]
\begin{subfigure}{.475\linewidth}
  \includegraphics[width=\linewidth]{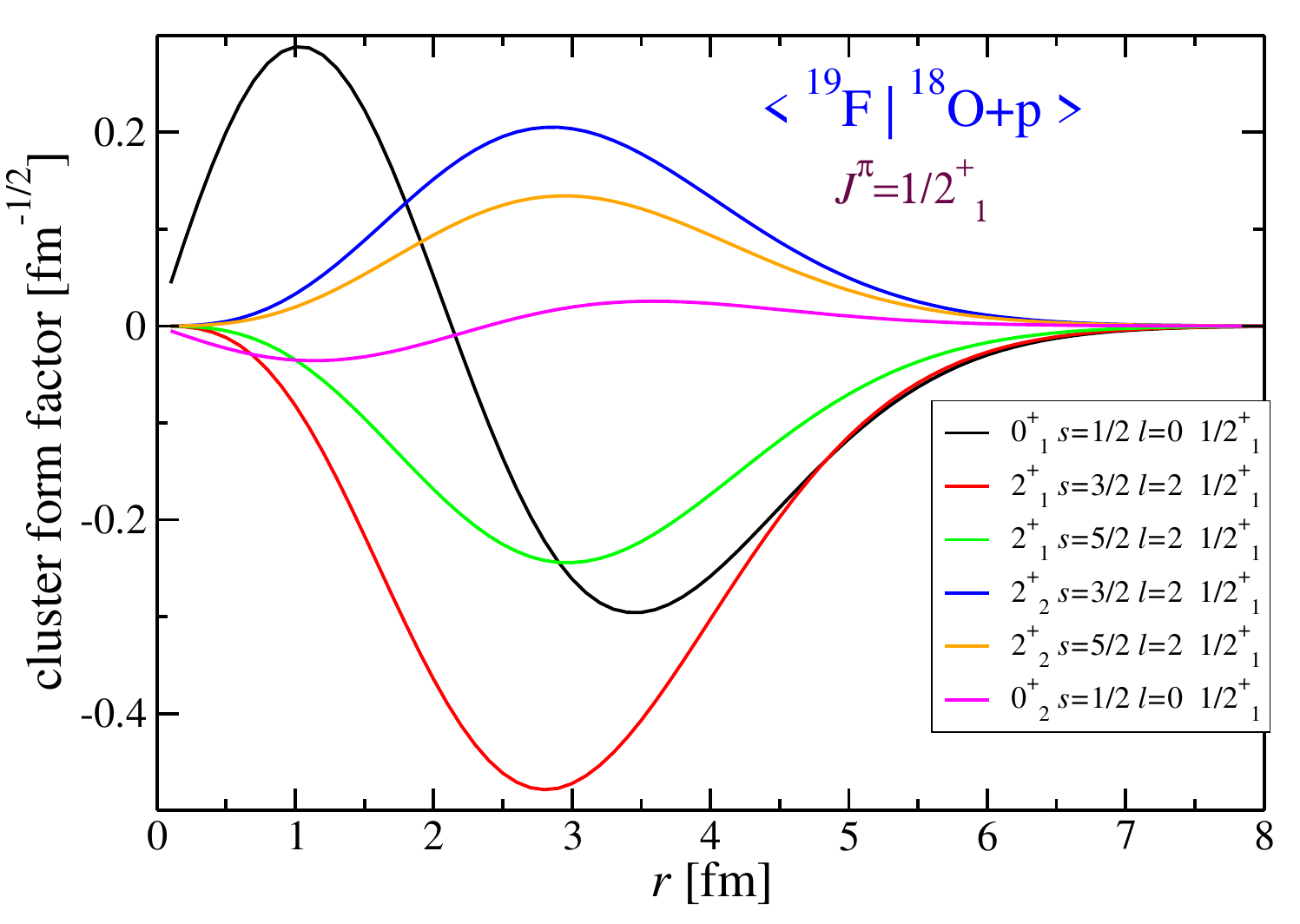}
  \caption{}
  \label{1}
\end{subfigure}\hfill 
\begin{subfigure}{.475\linewidth}
  \includegraphics[width=\linewidth]{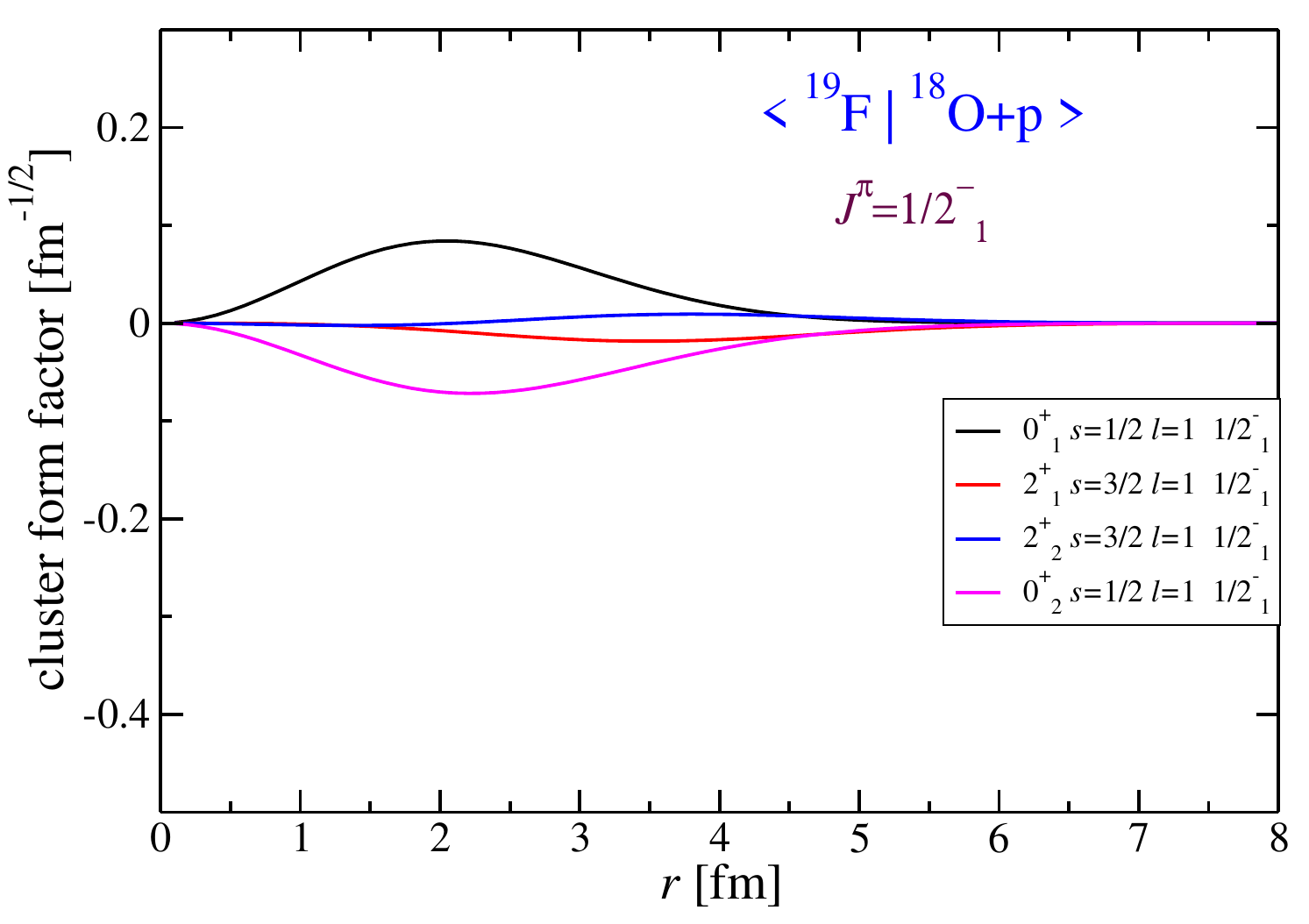}
  \caption{}
  \label{2}
\end{subfigure}
\medskip 
\begin{subfigure}{.475\linewidth}
  \includegraphics[width=\linewidth]{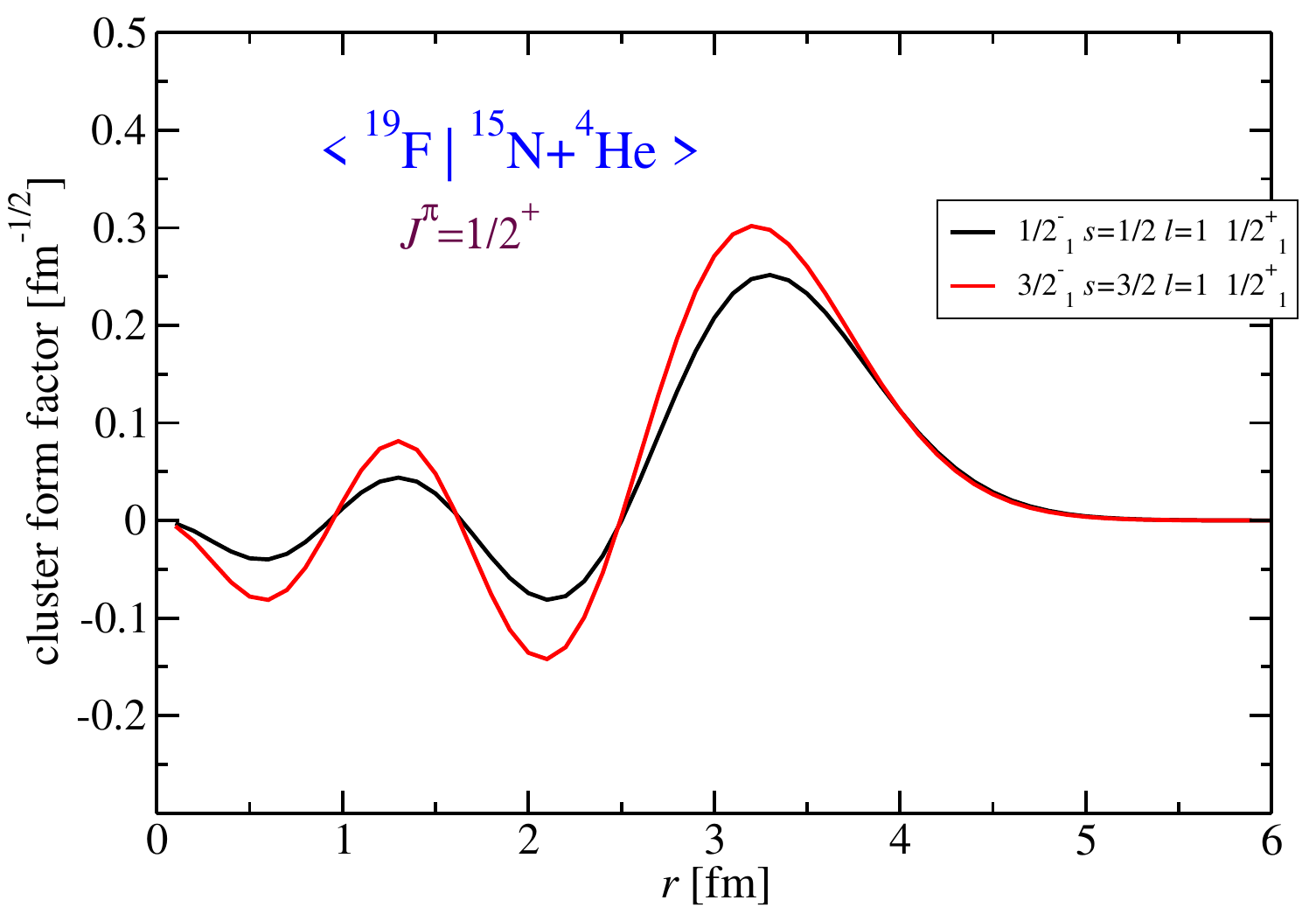}
  \caption{}
  \label{3}
\end{subfigure}\hfill 
\begin{subfigure}{.475\linewidth}
  \includegraphics[width=\linewidth]{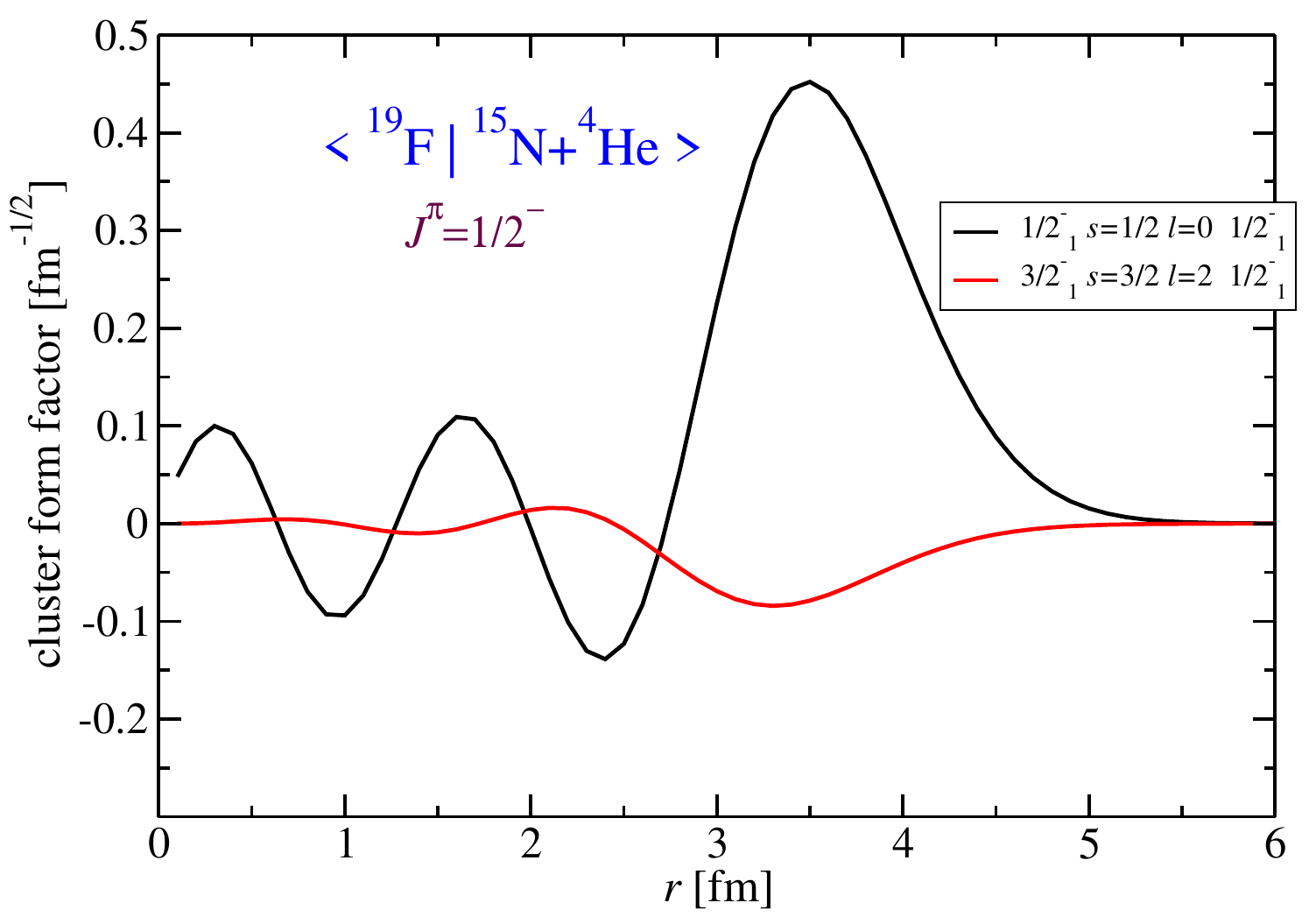}
  \caption{}
  \label{4}
\end{subfigure}
\caption{Cluster structure of $^{19}$F $1/2^+$ ground state (panels (a) and (c)) and the first excited $1/2^-$ state (panels (b) and (d)).The wave functions were obtained using the SRG-evolved NN-N$^4$LO + 3N$^*_\mathrm{lnl}$ interaction in $N_{\rm max}{=}4{-}6$ basis spaces and the HO frequency of $\hbar\Omega{=}18$ MeV. See the text for further details.}
\label{fig:overlaps}
\end{figure}
As seen in Fig.~\ref{fig:PV_moments_full-vs-1m-contr}, the NSM of $^{19}$F is dominated by the contribution of low-lying $1/2^-_1$ state. This is not the case for the nuclear EDM of $^{19}$F where the contribution of the lowest opposite-parity partner of the ground-state is negligible. The reason for that is a differing structure of the two states revealed in Fig.~\ref{fig:overlaps} where cluster form factors~\cite{Navratil2004} of the two states are shown considering $^{18}$O+p and $^{15}$N+$\alpha$ clusters. The $1/2^+$ ground state is a shell-model like state with large $S$-wave $^{18}$O(g.s.)+p and $D$-wave $^{18}$O($2^+_1$)+p amplitudes (Fig.~\ref{fig:overlaps} (a)) with small $\alpha$ cluster amplitudes dominated by the $^{15}$N($3/2^-_1$)+$\alpha$ $P$-wave (Fig.~\ref{fig:overlaps} (c)). On the contrary, the $1/2^-_1$ state exhibits $\alpha$-clustering with large $S$-wave $^{15}$N(${\rm g.s.},1/2^-_1$)+$\alpha$ amplitude ((Fig.~\ref{fig:overlaps} (d)) and a negligible $P$-wave $^{18}$O+p amplitudes (Fig.~\ref{fig:overlaps} (b)). Consequently, the matrix elements of the $E1$ operator and of the second term of the Schiff operator in Eq.~(2) in the main text ($\varpropto \mathbf{r}$) are very small while that of the first term of the Schiff operator ($\varpropto r^2 \mathbf{r}$) is enhanced.

In addition to the one-pion-exchange contributions to the $^{19}$F NSM, we also computed the contributions due to the $\rho-$ and $\omega-$meson exchanges~\cite{Liu2004} whose magnitude we find a factor of 5 to 10 smaller. These are related to the contact term in $V^{\rm PTV}$~\cite{deVries2020,engel2025nuclear} that we are in the process of implementing. Importantly, for CP-violating sources that preserve chiral symmetry, the corresponding contact interactions enter at the same chiral order as the pion-exchange contribution and can therefore play a quantitatively significant role~\cite{deVries2020}. We are currently implementing these operators in our framework. We have also initial results for the nucleon contributions to the NSM~\cite{engel2025nuclear}. We will provide further details on the NSM calculations for $^{19}$F and other, lighter, nuclei in Ref.~\cite{Foster2025}.

\clearpage
\section{Molecular calculations}

We have performed a thorough analysis of the remaining errors in the treatments of basis-set, electron-correlation, and relativistic effects in the quantum chemistry calculations for the $^{19}$F NSM molecular sensitivity factor in HfF$^+$. 

\begin{table}[b]
           \caption{\textbf{Computed molecular sensitivity factors to $^{19}$F nuclear Schiff moment in HfF$^+$.} $W_S$ is expressed in units of $\frac{e}{4\pi \epsilon_0 a_0^4} \approx 44.3~h~\mathrm{Hz}/(e~\mathrm{fm}^3)$.}
            \label{tab:WS}
            \begingroup
            \setlength{\tabcolsep}{4pt}
            \centering
            \rowcolors{2}{gray!20}{white}
            \begin{tabular}{c c c c c}
                \toprule
F basis sets	     &        Hf basis sets	  & frozen core electrons &	HF	& CCSD \\
   \midrule
ETB0 (30s30p4d3f2g)	  &  ANO-RCC(24s21p15d11f4g2h) &	48	& 122.0	& 117.5 \\
ETBSPD (30s30p30d3f2g) &	ANO-RCC(24s21p15d11f4g2h)&	48	& 122.1	& 117.5 \\
ETBSP1.8 (51s51p4d3f2g)	& ANO-RCC(24s21p15d11f4g2h)   & 48	& 122.1	& 117.6 \\
 \hline
ETBTZ(30s30p3d2f)	&    DYVTZ(30s24p15d11f2g)	     &  48	& 122.3 & 116.4 \\
ETBQZ(30s30p4d3f2g)	&    DYVQZ(34s30p19d13f4g2h)	 &  48	& 122.1	& 117.5 \\
ETBTZ(30s30p3d2f)	&    DYVTZ(30s24p15d11f2g)	     &  28	& 122.3	& 116.2 \\
ETBQZ(30s30p4d3f2g)	&    DYVQZ(34s30p19d13f4g2h)	 &  28	& 122.1	& 117.4 \\
                \bottomrule
            \end{tabular}
            \endgroup
            \label{tabbas}
        \end{table}

1) Basis-set effects. The present calculations have used extensive basis sets for F (30s30p4d3f2g) and Hf (24s21p15d11f4g2h). The F basis set adopts the primitive functions of F and includes many additional s- and p-type tight functions obtained using a geometric factor of 2.5. The Hf basis set is obtained by decontracting the ANO-RCC set of Hf. These basis sets are of quadruple-zeta quality for the treatment of electron correlation for both F and Hf. Further addition of steep functions for F has negligible effects. As shown in Table \ref{tabbas}, further inclusion of tight d-type functions for F (30s30p30d3f2g) only changes the computed sensitivity factor by less than 0.1 a.u. (less than 0.1\%). The use of a denser set of s- and p-type functions (51s51p4d3f2g) leads to a small change of 0.1 a.u. 

The use of the unc-ANO-RCC set for Hf provides results in close agreement with those obtained by using Dyall's quadruple-zeta (QZ) basis set for Hf, with discrepancies less than 0.1 a.u. The difference between the result obtained using the QZ basis and that using triple-zeta (TZ) basis amounts to around 1 a.u. We performed a basis set extrapolation and estimated a remaining effect of around 0.8 a.u. (less than 1 \%). Therefore, the remaining basis set correction is expected to be around 1 a.u. (1\% of total value). 

2) Electron correlation effects. The X2CAMF-CCSD and CCSD(T) calculations have correlated the F 2s2p and Hf 5s5p4f6s5d electrons together with virtual spinors below 100 Hartree. The total CCSD contribution is around -5 a.u. (the difference between the CCSD value of 117.5 a.u. and the HF value of 122.2 a.u. in Table \ref{tabbas}), which is around 4\% of the total value. The (T) contribution amounts to -2 a.u. (the CCSD(T) value of 115.3 a.u. in Table I of the main text and the CCSD value of 117.5 a.u. of Table \ref{tabbas}), which is around 2\% of the total value. As a conservative estimate, we expect that the high-level correlation effects beyond CCSD(T) is less than the CCSD contribution. The core-correlation effects are negligible. They have been studied by further correlating the F 1s and Hf 4s4p4d electrons as well as virtual spinors below 1000 Hartree. This core-correlation contribution is found to be less than 0.1 a.u. (less than 0.1\%). Therefore, we conclude that the remaining electron-correlation contribution is less than 4\% of the total value.

3) We performed a four-component Dirac-Coulomb-Gaunt (DCG) Hartree-Fock calculation using TZ basis and found that the difference between four-component DCG and X2CAMF(DCG) Hartree-Fock results is around 0.1 a.u. (less than 0.1\%). The remaining relativistic correction thus is insignificant. 

In summary, the analysis of the remaining errors in the quantum chemistry calculations concludes that the uncertainty of the computed $^{19}$F NSM molecular sensitivity factor is conservatively estimated as below 10\% of the total value. 

\bibliography{biblio}